\documentclass[12pt,draftcls, onecolumn]{IEEEtran}
\usepackage{graphicx}
\usepackage{amsmath}
\usepackage{cite}
\usepackage{amssymb}
\usepackage{multirow}
\usepackage{color}
\usepackage{epstopdf}
\usepackage{float}
\usepackage{algorithm}
\usepackage{algpseudocode}
\usepackage{amsmath,amssymb}
\usepackage{flushend}

\makeatletter
\let\MYcaption\@makecaption
\makeatother

\usepackage[font=footnotesize]{subcaption}

\makeatletter
\let\@makecaption\MYcaption
\makeatother

\usepackage{graphicx}
\usepackage{bm}
\usepackage{tabularray}

 \IEEEoverridecommandlockouts
 \def\BibTeX{{\rm B\kern-.05em{\sc i\kern-.025em b}\kern-.08em
 		T\kern-.1667em\lower.7ex\hbox{E}\kern-.125emX}}
\begin{document}

%
\title{Joint Channel Estimation and Localization in Pinching-Antenna OFDM Systems: The Blessing of Multipath}
\author
{
	Min Liu, Yue Xiao, Shuaixin Yang, Gang Wu, Xianfu Lei and Wei Xiang~\IEEEmembership{}
	
	 ~\IEEEmembership{}

	\thanks{
		M. Liu, Y. Xiao, S. Yang and G. Wu are with the National Key Laboratory of Wireless
		Communications, University of Electronic Science and Technology of
		China (UESTC), Chengdu 611731, China (e-mail: liumin\_uestc@std.uestc.edu.cn, xiaoyue@uestc.edu.cn, shuaixin.yang@foxmail.com,  wugang99@uestc.edu.cn).
		
		X. Lei is with the School of Information Science and Technology, Southwest Jiaotong University, Chengdu 610031, China (e-mail:
		xflei@swjtu.edu.cn).
		
		W. Xiang is with the School of Engineering and Mathematical Sciences, La Trobe University, Victoria 3086, Australia (e-mail: W.Xiang@latrobe.edu.au).
		
	}
}
\maketitle
\begin{abstract}

Pinching-antenna systems (PASS) have recently attracted considerable attention owing to their capability of flexibly reconfiguring large-scale wireless channels. 
Motivated by this potential, we investigate the issue of joint localization and channel estimation for the uplink PASS in the presence of multipath dispersion. 
To this end, a comprehensive multi-user orthogonal frequency division multiplexing (OFDM) uplink PASS model is first established, where the use of a cyclic prefix (CP) enables the multipath-induced time-domain dispersion to be transformed into a set of superimposed sinusoids in the frequency domain. 
Building upon this model, we propose a hybrid inference framework capable of accurately estimating both channel parameters and user locations. Specifically, expectation propagation is first employed to mitigate multi-user interference, while the path delays are then extracted from noisy channel state information using an orthogonal matching pursuit (OMP) based approach, or a hybrid belief propagation-variational inference (BP-VI) algorithm. 
Then the estimated delays are subsequently refined through the embedded geometric information via an iterative localization procedure, wherein the estimated channel matrices are recursively fed back to EP. 
Furthermore, the Cramér–Rao lower bound (CRLB) is derived to characterize the fundamental estimation limits. Finally, simulation results validate that our proposed framework closely approaches the CRLB, with performance comparable to cooperative multi-base station localization, with significantly fewer RF chains and reduced hardware complexity.

\end{abstract}


\IEEEpeerreviewmaketitle

\section{Introduction}

The forthcoming generation of wireless communication systems\cite{6G1,6G2} is expected to deliver multi-gigabit-per-second data rates, which in turn imposes stringent constraints on radio frequency (RF) front-end hardware and antenna architecture design. Nevertheless, wireless fading in both small scale and large scale remains a formidable challenge, impeding the reliability, coverage, and quality-of-service (QoS) guarantees in practical deployments. To overcome this challenge and to enhance the effective signal-to-noise ratio (SNR), flexibly positioned antenna architectures have emerged as a promising solution and attracted increasing research interest. In contrast to conventional fixed-position antennas, reconfigurable systems such as fluid antenna systems (FAS) \cite{FA1,MA1} and pinching antennas \cite{PA1,PA2,PA3,PAloss} (PAs) dynamically adapt the the physical antenna location. As a result, they are capable of mitigating both small-scale multipath fading, thereby substantially enhancing the received SNR at the user equipment (UE) and unlocking significant performance gains in terms of link reliability and spectral efficiency.
Specifically, as modeled in \cite{FA1,MA1}, fluid or movable antenna systems exploit position-flexible antenna elements capable of dynamically relocating within a confined region to harness spatial diversity and mitigate small-scale fading. This spatial agility has catalyzed a broad spectrum of applications, encompassing adaptive beamforming \cite{FABF1,FABF2}, integrated sensing and communication (ISAC) \cite{FAISAC,FAISAC2}, physical-layer security \cite{FAPLS}, and index modulation \cite{FAIM,FACC}. In tandem with these innovations, the development of accurate channel estimation techniques tailored for such flexible architectures has also been actively explored \cite{FACE1,MACE1}. In all, FAS has been shown to yield considerable performance gains over conventional architectures.

Recently, pinching-antenna systems (PASS) have also emerged as a particularly compelling architecture. In PASS deployments, electromagnetic (EM) waves are guided via a dielectric waveguide, while the actual radiation is facilitated by multiple reconfigurable dielectric particles, which are also termed as pinching antennas. Unlike fluid antennas, whose mobility is inherently confined to a small spatial aperture, the dielectric waveguide in PASS can be engineered to span extended spatial domains with marginal in-waveguide attenuation. This facilitates reconfigurable PA placements over a substantially larger region, thereby endowing the system with a degree of spatial freedom unattainable by conventional movable-antenna designs.

Owing to these structural advantages, each user equipment (UE) can establish communication with one or more pinching antennas via short-range line-of-sight (LoS) paths, thereby significantly mitigating path loss and enabling advanced transmission strategies. For instance, substantial throughput and energy-efficiency gains have been demonstrated for rate optimization \cite{PARate,PARate2} and beamforming \cite{PABF1,PABF2,PABFZY} in the context of PASS. The fundamental concepts and theoretical limits of PASS were initially characterized in \cite{PA1,PAloss}, where both single- and multi-waveguide configurations were analyzed in terms of channel capacity. In \cite{PARate,PARate2}, optimal PA placements were devised to minimize path loss and hence maximize downlink rates, including under non-orthogonal multiple access (NOMA) scenarios. Furthermore, hybrid beamforming designs subject to PA-location constraints were explored in \cite{PABF1}, showcasing the flexibility of PASS in spatial-domain control. Beyond throughput maximization, PASS has also been employed to enhance physical-layer security \cite{PAPLS1} by jointly configuring the amplitude and phase responses across different PAs to simultaneously ensure high-quality service for legitimate users while degrading the channel conditions of potential eavesdroppers. In addition, index modulation and directional modulation techniques have been successfully integrated into PASS \cite{PAPLSZY}, offering further enhancements in secure communications.

From a sensing and ISAC perspective, PASS has been predominantly studied through the Cramér–Rao lower bound (CRLB) analysis. In \cite{PASensing8}, the Bayesian CRLB was derived for a multi-target sensing scenario in a PASS-enabled uplink. The joint sensing–communication tradeoff was further quantified in \cite{PASensing5}, characterizing achievable rate regions in PASS-ISAC systems. The multi-target localization problem was revisited in \cite{PASensing1}, where PAs serve as probing nodes and leaky coaxial cables are employed as receivers, accompanied by CRLB analysis. In \cite{PASensing2}, a novel ISAC framework was proposed wherein the PAs act as active transmitters and a uniform linear array (ULA) captures the echoes, enabling joint PA location optimization via CRLB minimization. Furthermore,~\cite{PASensing3} investigated a two-waveguide ISAC system comprising a transmitting PA and a receiving PA, where the illumination power was optimized under QoS constraints, and the bistatic radar sensing CRLB was derived in~\cite{PASensing4}. Notably, \cite{PASensing6} and \cite{PASensing7} are among the few works investigating received-signal-strength (RSS)-based localization. In \cite{PASensing6}, PASS downlink RSS was leveraged under a Poisson line process waveguide distribution model, whereas \cite{PASensing7} considered indoor uplink localization via a weighted least squares (WLS) approach. In general, most existing PASS-based sensing and ISAC schemes emphasize angular information, signal amplitude, or phase manipulation, relatively limited attention has been paid to delay-domain signal processing.

In uplink PASS scenarios where multiple PAs on the same waveguide receive the sensing or communication signal, the associated signal \emph{delays} at different PAs are generally distinct. This inherently gives rise to multipath-induced dispersion, potentially resulting in severe intersymbol interference (ISI) \cite{PAOFDM}. The underlying cause stems from the varying propagation delays of individual UE–PA links. At low sampling rates, i.e., when the symbol duration significantly exceeds the maximum excess delay, such dispersion is typically negligible. In contrast, at high sampling rates, the delay spread becomes pronounced and must be explicitly addressed. Specifically, \cite{PAOFDM} modeled the frequency-selective characteristics of PASS with multiple simultaneously active PAs using a finite impulse response (FIR) filter, while orthogonal frequency-division multiplexing (OFDM) with a cyclic prefix (CP) was adopted in \cite{PAOFDM,PAOFDM2} to mitigate the resulting multipath effects.

Nevertheless, it is important to note that multipath propagation is not always detrimental from a sensing perspective. In fact, the path delay corresponding to a specific UE–PA pair is directly related to their geometric distance via the speed of light. With a sufficient number of spatially distributed PAs, the UE position can be inferred by estimating these delays and exploiting the underlying geometric constraints. Compared with RSS-based metrics, propagation delays tend to be more stable, as they are less sensitive to fading \cite{Loc1}. By appending a sufficiently long CP to the start of each OFDM frame, the superimposed delayed replicas of the transmitted signal are converted into cyclic time-domain shifts, which manifest as phase-rotated sinusoids in the frequency domain. By extracting these phase shifts, the corresponding delays and hence the geometric information can be recovered. Motivated by these observations, this paper investigates the joint channel estimation and localization problem in a multi-user uplink PASS employing OFDM, explicitly exploiting the inherent multipath dispersion in PASS.

The main contributions of this paper are then summarized as follows:
\begin{enumerate}

	\item \textbf{System Modeling:} We develop a detailed system model for multi-user uplink PASS employing OFDM. The distinct propagation delays associated with different PAs naturally induce delay dispersion. By incorporating the CP, we transform the time-domain mutipath channel to a circularly shift-variant convoluted version, corresponding to a set of superimposed phase-rotated sinusoids in the frequency domain. By aggregating multiple OFDM frames, we then derive a compact and tractable formulation of the multi-user transmission model.
	
	\item \textbf{Joint Channel Estimation and Localization Framework:} We propose a Joint Channel Estimation and Localization (JCEL) framework tailored to uplink PASS with OFDM. Specifically, an expectation propagation (EP)\cite{EP1,EP2,VAMP} based iterative receiver is designed to mitigate multi-user interference and decouple the underlying parameters. Based on the initial coarse channel estimates, the delay-extraction subproblem is tackled via two alternatives: a compressed sensing-based orthogonal matching pursuit (OMP) method, or a belief propagation and variational inference (BP-VI)\cite{VALSE,VMP} off-grid estimator that jointly refines the delays and channel coefficients. Subsequently, a delay-based localization procedure is introduced, which iteratively combines delay matching with a soft position-refinement step to recover UE locations.

\item \textbf{Theoretical Bound and Performance Evaluation:} We derive the associated CRLB and conduct extensive numerical simulations. The results show that the proposed framework is capable of jointly estimating the uplink channel and accurately localizing UEs, achieving performance close to the CRLB and comparable to that of conventional multi-BS cooperative localization benchmarks, while significantly reducing the number of required RF chains due to the PASS architecture.
\end{enumerate}

\emph{Notation:} 
Bold lowercase and uppercase letters denote column vectors and matrices, respectively. 
The $(x,y)$-th element of matrix $\mathbf{A}$ is denoted by $[\mathbf{A}]_{x,y}$. 
The operators $(\cdot)^{H}$ and $(\cdot)^{T}$ represent the conjugate transpose and transpose, respectively, and $(\cdot)^{*}$ denotes the conjugate.
The notation $\mathcal{CN}(\mathbf{x}; \boldsymbol{\mu}, \mathbf{\Sigma})$ denotes the probability density function (PDF) of a circularly symmetric complex Gaussian distribution with mean vector $\boldsymbol{\mu}$ and covariance matrix $\mathbf{\Sigma}$. 
Furthermore, $\mathcal{VM}(x; \mu, \kappa)$ denotes the PDF of a Von~Mises distribution with mean direction $\mu$ and concentration parameter $\kappa$. 
The operator $\mathbb{E}(\cdot)$ denotes statistical expectation, and $\mathrm{Vec}(\cdot)$ denotes the vectorization operation of a matrix. $Tr\left(\cdot\right)$ denotes the trace of a matrix.

\section{Uplink PASS Model}

\begin{figure}[!t]
	\centering
	\includegraphics[width=\linewidth]{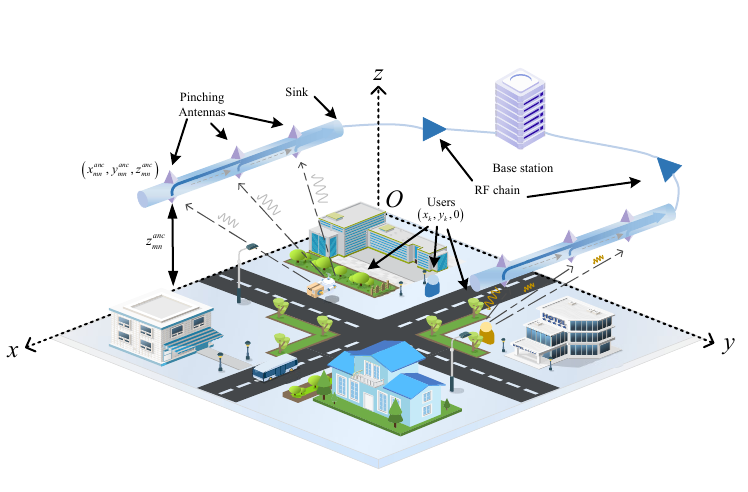}
	\caption{PASS uplink receiving model with $M = 2$, $N_1 = N_2 = 3$, and $K = 3$.}
	\label{fig:Sysmodel}
\end{figure}

In a pinching-antenna system (PASS), each pinching antenna (PA) is realized by a small dielectric particle attached to a segment of a dielectric waveguide. 
In the uplink scenario, the EM wave radiated by the users is received by the PAs and then guided along the dielectric waveguides towards the RF chains at the base station, as illustrated in Fig.~\ref{fig:Sysmodel}. 
We consider a receiver architecture employing multiple waveguides, each equipped with multiple PAs and terminates at independent sinks where the signals are sampled.

Specifically, assume that $M$ waveguides are deployed at a common fixed height $z^{\mathrm{anc}}$, and that the $m$-th waveguide accommodates $N_m$ PAs. 
The total number of PAs is thus 
\(
N_{\mathrm{all}} = \sum_{m=1}^{M} N_m.
\)
The three-dimensional coordinate of the sink of the $m$-th waveguide is denoted by
\(
\boldsymbol{\psi}_m^{\mathrm{anc}} = \left( x_m^{\mathrm{anc}},\, y_m^{\mathrm{anc}},\, z^{\mathrm{anc}} \right),
\)
while the coordinate of the $n$-th PA on the $m$-th waveguide is written as
\(
\boldsymbol{\psi}_{mn}^{\mathrm{anc}} = \left( x_{mn}^{\mathrm{anc}},\, y_{mn}^{\mathrm{anc}},\, z^{\mathrm{anc}} \right).
\)

We consider $K$ legitimate users located on the ground plane. 
The coordinate of the $k$-th user is
\(
\boldsymbol{\psi}_k = \left( x_k,\, y_k,\, 0 \right),
\)
with
\(
x_k \in [C_x^{\min}, C_x^{\max}]
\)
and
\(
y_k \in [C_y^{\min}, C_y^{\max}].
\)
The baseband channel impulse response from the $k$-th user to the $(m,n)$-th PA is modeled as
\begin{equation}
	h_{kmn}^{o}(t)
	= z_{kmn}^{o}
	\,\delta\!\left( t - \tau_{kmn}^{o} \right),
\end{equation}
where $z_{kmn}^{o}$ denotes the complex fading coefficient. For free-space fading, we have 
$z_{kmn}^o = \frac{{{\eta ^{\frac{1}{2}}}}}{{\left\| {{\bm\psi _k} - \bm\psi _{mn}^\mathrm{anc}} \right\|}}e^{j\phi_{kmn}^o}$
where
\(
\eta = \frac{c^2}{16 \pi^2 f_c^2}
\)
denoting the free-space reference loss at $1$~m, with $f_c$ and $c$ being the carrier frequency and the speed of light in vacuum, respectively. Furthermore, $\phi_{kmn}^o= 2\pi \frac{{{f_c}\left\| {{\psi _k} - \psi _{mn}^\mathrm{anc}} \right\|}}{c}$ is the phase rotation of $z_{kmn}^o$ induced by propagation. For other fading types, the structure of $z_{kmn}^{o}$ varies, and
\(
\tau_{kmn}^{o}
= \frac{\left\| \boldsymbol{\psi}_k - \boldsymbol{\psi}_{mn}^{\mathrm{anc}} \right\|}{c}
\)
is the corresponding free-space propagation delay.

The equivalent channel between the $(m,n)$-th PA and the sink of the $m$-th waveguide is modeled as
\begin{equation}
	h_{mn}^{i}(t)
	= z_{mn}^{i} \,\delta\!\left( t - \tau_{mn}^{i} \right),
\end{equation}
where
\(
\tau_{mn}^{i}
= \frac{\left\| \boldsymbol{\psi}_m^{\mathrm{anc}} - \boldsymbol{\psi}_{mn}^{\mathrm{anc}} \right\|}
{n_d c}
\)
is the in-waveguide propagation delay, $n_d$ is the effective refractive index of the dielectric waveguide, and $z_{mn}^{i}$ is the complex in-waveguide fading coefficient capturing attenuation and additional phase shifts between the $(m,n)$-th PA and the $m$-th sink.

\begin{figure}[htbp]
	\centering
	\includegraphics[width=0.7\linewidth]{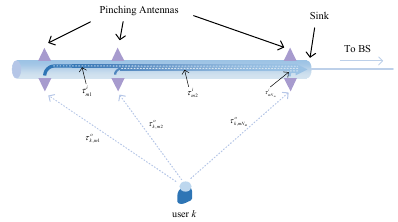}
	\caption{Propagation model for the $k$-th user and the $(m,n)$-th PA, illustrating the in-waveguide and free-space propagation delays.}
	\label{fig:Sysmodel2}
\end{figure}

By cascading the free-space and in-waveguide channels, the overall impulse response from the $k$-th user to the $m$-th sink is given by
\begin{equation}
	h_{km}(t)
	= \sum_{n=1}^{N_m} h_{kmn}^{o}(t) * h_{mn}^{i}(t)
	= \sum_{n=1}^{N_m} z_{kmn} \,\delta\!\left( t - \tau_{kmn} \right),
\end{equation}
where
$z_{kmn}=z_{mn}^{i} z_{kmn}^{o}$
is the composite complex fading coefficient incorporating both free-space and in-waveguide effects, and
\(
\tau_{kmn} = \tau_{kmn}^{o} + \tau_{mn}^{i}
\)
denotes the total propagation delay along the corresponding path. 
The underlying propagation geometry is depicted in Fig.~\ref{fig:Sysmodel2}.

Since $\tau_{mn}^{i}$ is determined solely by the fixed waveguide geometry and can be calibrated offline, the composite delay $\tau_{kmn}$ is uniquely determined by the physical distance between the $k$-th user and the $(m,n)$-th PA. 
Hence, the spatially separated PAs may act as effective virtual anchors for localization, provided that the individual delay can be reliably extracted and associated with the corresponding PAs.

We assume that the $K$ users transmit pilot symbols for channel estimation and localization. 
Let $x_{kp}(t)$, $t \in [0, LT]$, denote the pilot waveform of the $k$-th user during the $p$-th OFDM block, where $T$ and $L$ represent the sampling interval and the number of useful time-domain samples per block, respectively. 
The received signal at the $m$-th waveguide can then be expressed as
\begin{equation}
	y_{mp}(t)
	= \sum_{k=1}^{K} \bigl( h_{km}(t) * x_{kp}(t) \bigr)
	+ n_m(t),
\end{equation}
where $n_m(t)$ denotes the additive white Gaussian noise (AWGN) at the $m$-th sink.

By appending a cyclic prefix (CP) with length $L_{\mathrm{CP}} T$ to each pilot block, with
\(
L_{\mathrm{CP}} T \ge \max_{k,m,n} \tau_{kmn},
\)
and sampling the received signal with period $T$, the resulting discrete-time model for the $k$-th user, $m$-th waveguide, and $p$-th block becomes
\begin{equation}
	\begin{aligned}
		y_{kmp}[i]
		&= \mathrm{Circonv}\bigl( h_{km}[i],\, x_{kp}[i] \bigr) + n_{kmp}[i] \\
		&= \sum_{n=1}^{N_m} 
		z_{kmn}\,
		\mathrm{Cirshift}\bigl( x_{kp}[i];\, \tau_{kmn} \bigr)
		+ n_{kmp}[i],
	\end{aligned}
\end{equation}
for $i = 0, \ldots, L-1$, where $\mathrm{Circonv}(\cdot,\cdot)$ and $\mathrm{Cirshift}(\cdot;\cdot)$ denote circular convolution and circular time shifting, respectively, both with implicit periodic extension.

Applying an $L$-point discrete Fourier transform (DFT) to the time-domain samples yields the frequency-domain model
\begin{equation}
	Y_{kmp}[l]
	= H_{km}[l]\, X_{kp}[l]
	+ n_{kmp}[l], \qquad l = 0,\ldots,L-1,
\end{equation}
where $X_{kp}[l]$ and $Y_{kmp}[l]$ are the DFTs of $x_{kp}[i]$ and $y_{kmp}[i]$, respectively, and
\begin{equation}\label{H}
	H_{km}[l]
	= \sum_{n=1}^{N_m}
	z_{kmn}
	\exp\!\left(
	- j 2 \pi \frac{\tau_{kmn}}{L T} \, l
	\right)
\end{equation}
is the equivalent frequency-domain channel coefficient of the $k$-th user towards the $m$-th waveguide.

Collecting the channels of all $K$ users towards all $M$ waveguides, the frequency-domain MIMO channel matrix at subcarrier $l$ is given by
\begin{equation}
	\mathbf{H}[l]
	= \begin{bmatrix}
		H_{11}[l] & \cdots & H_{K1}[l] \\
		\vdots     & \ddots & \vdots     \\
		H_{1M}[l] & \cdots & H_{KM}[l]
	\end{bmatrix}
	\in \mathbb{C}^{M \times K}.
\end{equation}

Assuming that each user transmits a constant pilot symbol within each frame and aggregating $P$ pilot frames, the pilot matrix $\mathbf{X} \in \mathbb{C}^{K \times P}$ is written as
\begin{equation}
	\mathbf{X}
	= \begin{bmatrix}
		\mathbf{x}_1^{T} \\
		\vdots \\
		\mathbf{x}_K^{T}
	\end{bmatrix},
	\qquad
	\mathbf{x}_k
	= \bigl[ x_{k1}, \ldots, x_{kP} \bigr],
\end{equation}
and the corresponding stacked observation model at subcarrier $l$ becomes
\begin{equation}\label{SystemModel}
	\mathbf{Y}[l]
	= \mathbf{H}[l] \mathbf{X}
	+ \mathbf{N}[l] \in \mathbb{C}^{M \times P},
\end{equation}
where $\mathbf{Y}[l]$ collects the received pilot symbols at all $M$ sinks across $P$ frames, and $\mathbf{N}[l]$ denotes the additive noise matrix whose entries are i.i.d. $\mathcal{CN}(0,\sigma_n^2)$.

\textit{Remark:} 
Unlike conventional cooperative localization systems, where each base station directly observes the user signal, the PASS receiver observes a superposition of phase-rotated sinusoids originating from multiple PAs on each waveguide. 
Consequently, in order to enable joint channel estimation and localization, the underlying path delays must first be extracted from the frequency-domain measurements and then reliably associated with their corresponding PAs.

\section{Iterative Channel Estimation}

As illustrated in~\eqref{SystemModel}, the received signal ${\bf Y}[l]$ at each subcarrier
is a superposition of the underlying channel responses ${\bf H}[l]$. 
Hence, it is imperative to recover ${\bf H}[l]$ from ${\bf Y}[l]$ for all subcarrier indices
$l = 1, \ldots, L$. 
By vectorizing ${\bf Y}[l]$ and stacking all $L$ subcarriers, the overall signal model can be expressed in the compact form
\begin{equation}
	\tilde{\bf y}
	=
	\tilde{\bf M}\,\tilde{\bf h}
	+
	\tilde{\bf n},
\end{equation}
where
$\tilde{\bf h} = {\rm Vec}\!\left(\left[{\bf H}[1], \ldots, {\bf H}[L]\right]\right)
\in \mathbb{C}^{MKL\times 1}$ denotes the vectorized composite channel,
$\tilde{\bf y} = {\rm Vec}\!\left(\left[{\bf Y}[1], \ldots, {\bf Y}[L]\right]\right)
\in \mathbb{C}^{MPL\times 1}$ represents the stacked received signal vector, and
$\tilde{\bf M} = {\bf I}_L \otimes \!\left({\bf X}^{T} \otimes {\bf I}_M\right)$
is the corresponding block-structured measurement matrix. 
The vector $\tilde{\bf n}$ collects the equivalent noise samples, with $\tilde{\bf n}={\rm Vec}\!\left(\left[{\bf N}[1], \ldots, {\bf N}[L]\right]\right)
\in \mathbb{C}^{MPL\times 1}$.

For the sake of employing a real-valued inference framework, we introduce the real-valued representations ${\bf y}=
\begin{bmatrix}
	\mathcal{R}(\tilde{\bf y})^{T} &
	\mathcal{I}(\tilde{\bf y})^{T}
\end{bmatrix}^{T}$, ${\bf h}=
\begin{bmatrix}
\mathcal{R}(\tilde{\bf h})^{T} &
\mathcal{I}(\tilde{\bf h})^{T}
\end{bmatrix}^{T},$ ${\bf n}=
\begin{bmatrix}
\mathcal{R}(\tilde{\bf n})^{T} &
\mathcal{I}(\tilde{\bf n})^{T}
\end{bmatrix}^{T}$, 
$\left(\sigma_{n}^r\right)^{2}=\frac{\sigma_{n}^{2}}{2}$, and define the equivalent real measurement matrix as
\begin{equation}
	{\bf M}=
	\begin{bmatrix}
		\mathcal{R}(\tilde{\bf M}) & -\mathcal{I}(\tilde{\bf M}) \\
		\mathcal{I}(\tilde{\bf M}) & \mathcal{R}(\tilde{\bf M})
	\end{bmatrix}.
\end{equation}
The system model then takes the real-valued linear form
\begin{equation}
	{\bf y} = {\bf M}{\bf h} + {\bf n},
\end{equation}
with the noise variance being $\left(\sigma_{n}^r\right)^{2}$, which will be used as the basis of the subsequent Bayesian inference procedure.

In order to estimate ${\bf h}$ from ${\bf y}$ while exploiting its internal structure, we invoke the expectation propagation (EP) framework~\cite{EP1,VAMP}. 
The exact posterior distribution of ${\bf h}$ is given by
\begin{equation}
	p({\bf h}\,|\,{\bf y})
	\propto
	\mathcal{N}\!\left({\bf y}; {\bf M}{\bf h}, \left(\sigma_{n}^r\right)^{2}{\bf I}_{MPL}\right)
	p({\bf h}),
\end{equation}
where $p({\bf h})$ denotes the (generally non-Gaussian) prior distribution of the real-valued channel vector. 
EP approximates this intractable posterior by replacing $p({\bf h})$ with a tractable Gaussian density, yielding
\begin{equation}
	q({\bf h}\,|\,{\bf y})
	\propto
	\mathcal{N}\!\left({\bf y}; {\bf M}{\bf h}, {\bf I}_{MPL}\right)
	q({\bf h}),
\end{equation}
with
\begin{equation}
	q({\bf h})
	=
	\mathcal{N}\!\left({\bf h}; \hat{\bf h}, \sigma_{h}^{2}{\bf I}_{MKL}\right)
\end{equation}
serving as an approximate marginal prior.

The EP iterations commence with the initialization
$\big(\sigma_{h}^{(i)}\big)^{2}=1$ and
$\hat{\bf h}^{(i)} = {\bf 0}_{MKL}$ at iteration $i=1$. 
In the $i$-th iteration, the posterior covariance and mean of the Gaussian approximation are updated as
\begin{equation}
	\label{sigmao}
	\big({\boldsymbol{\Sigma}}_{q}^{(i)}\big)^{-1}
	=
	\left(\sigma_{n}^{r}\right)^{-2}{\bf M}^{H}{\bf M}
	+
	\big(\sigma_{h}^{(i)}\big)^{-2}{\bf I}_{MKL},
\end{equation}
\begin{equation}
	\label{muo}
	{\boldsymbol{\mu}}_{q}^{(i)}
	=
	{\boldsymbol{\Sigma}}_{q}^{(i)}
	\left(
	\left(\sigma_{n}^{r}\right)^{-2}{\bf M}^{H}{\bf y}
	+
	\big(\sigma_{h}^{(i)}\big)^{-2}\hat{\bf h}^{(i)}
	\right),
\end{equation}
and the corresponding average variance is obtained as
\begin{equation}
	\big(\sigma_{q}^{(i)}\big)^{2}
	=
	\frac{1}{MKL}\,{\rm tr}\!\left({\boldsymbol{\Sigma}}_{q}^{(i)}\right).
\end{equation}
By subtracting the incoming information, the parameters of the Gaussian message passed from the linear EP module to the denoising module are given by
\begin{equation}
	\big(\sigma_{h}^{\text{in}(i)}\big)^{2}
	=
	\left(
	\big(\sigma_{q}^{(i)}\big)^{-2}
	-
	\big(\sigma_{h}^{(i)}\big)^{-2}
	\right)^{-1},
\end{equation}
\begin{equation}
	\hat{\bf h}^{\text{in}(i)}
	=
	\big(\sigma_{h}^{\text{in}(i)}\big)^{2}
	\left(
	\big(\sigma_{q}^{(i)}\big)^{-2}{\boldsymbol{\mu}}_{q}^{(i)}
	-
	\big(\sigma_{h}^{(i)}\big)^{-2}\hat{\bf h}^{(i)}
	\right).
\end{equation}

Given $\big(\sigma_{h}^{\text{in}(i)}\big)^{2}$ and $\hat{\bf h}^{\text{in}(i)}$, the denoising module exploits the structural prior of ${\bf h}$ and produces a Gaussian-approximated output of the form
\begin{equation}
	\hat q^{(i)}({\bf h})
	=
	\mathcal{N}\!\left(
	{\bf h};
	\hat{\bf h}^{\text{out}(i)},
	\big(\sigma_{h}^{\text{out}(i)}\big)^{2}{\bf I}_{MKL}
	\right),
\end{equation}
where $\hat{\bf h}^{\text{out}(i)}$ and $\big(\sigma_{h}^{\text{out}(i)}\big)^{2}$ denote the refined mean and variance, respectively. 
The updated EP parameters for the next iteration are then obtained as
\begin{equation}
	\big(\sigma_{h}^{(i+1)}\big)^{2}
	=
	\left(
	\big(\sigma_{h}^{\text{out}(i)}\big)^{-2}
	-
	\big(\sigma_{h}^{\text{in}(i)}\big)^{-2}
	\right)^{-1},
\end{equation}
\begin{equation}
	\hat{\bf h}^{(i+1)}
	=
	\big(\sigma_{h}^{(i+1)}\big)^{2}
	\left(
	\big(\sigma_{h}^{\text{out}(i)}\big)^{-2}\hat{\bf h}^{\text{out}(i)}
	-
	\big(\sigma_{h}^{\text{in}(i)}\big)^{-2}\hat{\bf h}^{\text{in}(i)}
	\right).
\end{equation}

\begin{figure}[!t]
	\centering
	\includegraphics[width=\linewidth]{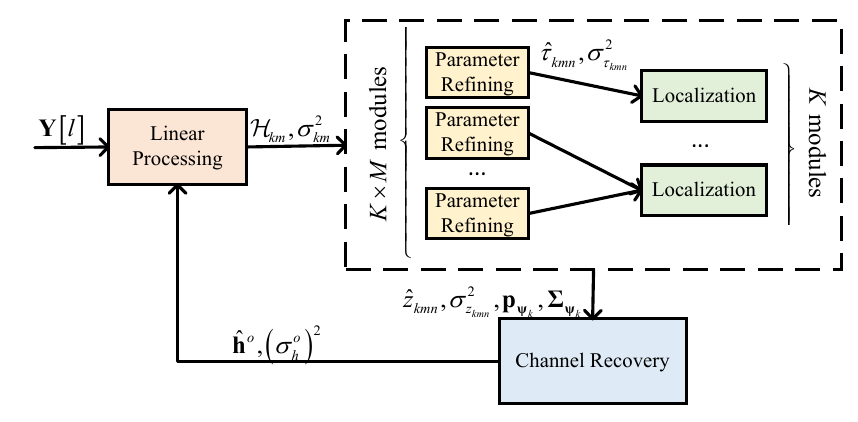}
	\caption{Block diagram of the EP-based iterative channel estimator.}
	\label{fig:IterDiagram}
\end{figure}

By iteratively refining $\hat{\bf h}$, $\sigma_{h}^{2}$, $\sigma_{q}^{2}$, and ${\boldsymbol{\mu}}_{q}$, the extended channel vector ${\bf h}$ is progressively improved, as summarized in Fig.~\ref{fig:IterDiagram}. 
To enhance numerical stability and avoid oscillations, a simple damping strategy can be employed as
\begin{equation}
	\hat{\bf h}_{\text{damp}}^{\text{in}(i)}
	=
	\gamma\,\hat{\bf h}^{\text{in}(i)}
	+
	(1-\gamma)\,\hat{\bf h}^{\text{in}(i-1)},
\end{equation}
\begin{equation}
	\big(\sigma_{h,\text{damp}}^{\text{in}(i)}\big)^{2}
	=
	\gamma\,\big(\sigma_{h}^{\text{in}(i)}\big)^{2}
	+
	(1-\gamma)\,\big(\sigma_{h}^{\text{in}(i-1)}\big)^{2},
\end{equation}
where $0 < \gamma \leqslant 1$ denotes the damping factor.

\section{Parameter Refinement}

In this section, the estimation of the intrinsic parameters, i.e., $\left(x_{k},y_{k}\right)$ and $z_{kmn}$, is refined and the parameters of the denoised result ${\hat q}\left({\bf h}\right)$ are calculated. This is equivalent to extracting the parameter with known noisy ${\bf H}[l]$. Specifically, we focus on the parameter extraction of the $k$-th user from the sink $m$. Hence, we first obtain the complex observation ${\mathcal{H}_{km}} \in {\mathcal C}^{L\times 1}$ by aggregating the $L$ slots of the noisy estimation of $H_{km}[l]$, in which ${\left[ {{\mathcal{H}_{km}}} \right]_l} = \left[ \hat{\bf h}^{\text{in}} \right]_{v_r}+j\left[ \hat{\bf h}^{\text{in}} \right]_{v_i}$ with ${v_r}={MK\left( {l - 1} \right) + \left( {k - 1} \right)M + m}$, $v_i=v_r+MKL$ and $\sigma _{km}^2 = 2{\left( {\sigma _h^{\text{in}}} \right)^2}$, where the iteration indicator $i$ is ignored for symbolic simplicity.

The position of the $k$-th user could be refined by performing exhausting search for the coordinate $\left(x_k,y_k,0\right)$. However, for higher dimensional systems, the corresponding complexity will be prohibitive. Hence, we turn to the acquisition of the delays $\{\tau_{kmn}\}$ and extract the localization information afterwards.

\subsection{Dictionary Based Delay Extraction Method}

To circumvent the need for a non-linear estimation procedure, the delay parameters 
$\{\tau_{kmn}\}$ could be discretized within the range of $[0, L_{\mathrm{CP}} T]$ into a finite 
set of candidate values, forming a delay dictionary. By constructing an overcomplete 
dictionary of size $N_d$ for $\{\tau_{kmn}\}$, the observation model corresponding to 
the $k$-th user at the $m$-th sink can be expressed as
\begin{equation}
	\label{SparseConstruct}
	{\mathcal H}_{km} 
	= 
	{\mathbf A}_{km}^{d} {\mathbf z}_{km}^{d} + {\mathbf n}_{km} 
	\in \mathbb{C}^{L \times 1},
\end{equation}
for $k = 1, \ldots, K$ and $m = 1, \ldots, M$, where 
${\mathcal H}_{km} = [H_{km}[0], \ldots, H_{km}[L-1]]^{T}$ 
denotes the reshaped observation vector at the $m$-th sink. 
The entries of the delay dictionary ${\mathbf A}_{km}^{d}$ are defined as $[{\mathbf A}_{km}^{d}]_{l', n_d} 
= 
\exp\!\left(
-j 2\pi \frac{L_{\mathrm{CP}} (n_d - 1)}{L N_d} (l' - 1)
\right)
$, $ n_d = 1, \ldots, N_d$, 
where $N_d$ serves as the grid-size parameter that provides a trade-off between the 
estimation accuracy and computational complexity. The coefficient vector 
${\mathbf z}_{km}^{d}$ is assumed to be $N_m$-sparse, while 
${\mathbf n}_{km}$ represents the equivalent additive white Gaussian noise (AWGN) 
with variance $\sigma_{km}^2$.

The sparse recovery problem formulated in~\eqref{SparseConstruct} can be efficiently 
solved by the well-established orthogonal matching pursuit (OMP) algorithm, as 
outlined in Algorithm~\ref{OMPalg}. The resulting delay estimates 
$\{\hat \tau_{kmn}\}$ and complex gain estimates $\{\hat z_{kmn}\}$ 
can subsequently be utilized for localization or signal reconstruction in the 
post-processing stage. 

\begin{algorithm}[t]
	\caption{OMP-Based Delay Extraction Procedure for ${\mathcal{H}_{km}}$}
	\begin{algorithmic}[1]
		\Require 
		Observation vector ${\mathcal H}_{km}$, dictionary ${\mathbf A}_{km}^{d}$, sparsity level $N_m$
		\Ensure 
		Estimated sparse vector ${\mathbf{\hat z}}_{km}^{d} \in \mathbb{C}^{N_d}$
		\State Initialize the residual ${\mathbf r} = {\mathcal H}_{km}$, 
		support set ${\mathcal S} \leftarrow \varnothing$, 
		and ${\mathbf{\hat z}}_{km}^{d} = \mathbf{0}_n$
		\For{$t = 1, 2, \ldots, N_m$}
		\State ${\mathbf c} = ({\mathbf A}_{km}^{d})^{H} {\mathbf r}$;
		\State $j_t = \arg \max_{i \notin {\mathcal S}} |{\mathbf c}_i|$;
		\State ${\mathcal S} \leftarrow {\mathcal S} \cup \{j_t\}$;
		\State ${\mathbf A}_{\mathcal S} = [{\mathbf A}_{km}^{d}]_{:,{\mathcal S}}$;
		\State ${\mathbf{\hat z}}_{km}^{d} = {\mathbf A}_{\mathcal S}^{\dagger} {\mathcal H}_{km}$;
		\State ${\mathbf r} = {\mathcal H}_{km} - {\mathbf A}_{\mathcal S} {\mathbf{\hat z}}_{km}^{d}$;
		\State $\hat \tau_{kmt} = j_t T L_{\mathrm{CP}} / N_d$;
		\State $\hat z_{kmt} = [{\mathbf{\hat z}}_{km}^{d}]_{j_t}$;
		\EndFor
		\State \Return ${\mathbf{\hat z}}_{km}^{d}$, $\{\hat \tau_{kmn}, \hat z_{kmn}\}_{n=1}^{N_m}$.
	\end{algorithmic}
	\label{OMPalg}
\end{algorithm}

\subsection{Message-Passing Based Delay Extraction Method}

Though the dictionary based method is intuitive, its precision is limited by the grid density. Thus, increasing the density of the grid may increase the correlation of the columns, as well as the algorithm complexity. Therefore, we propose a gridless hybrid message-passing based delay extraction method. In this section, for convenience, we use $\theta_{kmn}=-2\pi\frac{\tau_{kmn}}{LT}$, the equivalent angle, to replace $\tau_{kmn}$.

We first reformulate the observation of the $k$-th user at the $m$-th sink into parametric form as 
\begin{equation}
	\begin{aligned}
		{\mathcal{H}_{km}} = {\mathbf{A}}_{km}\left( {\boldsymbol{\theta }}_{km} \right){\mathbf{z}}_{km} + {\mathbf{n}}_{km}\\
		= \sum\limits_{n = 1}^{{N_m}} {{\mathbf{a}}\left( {{\theta _{kmn}}} \right){z_{kmn}}}  + {{\mathbf{n}}_{km}},
	\end{aligned}
\end{equation}
where ${{\mathbf{A}}_{km}}\left( {{{\boldsymbol{\theta }}_{km}}} \right) = \left[ {{\mathbf{a}}\left( {{\theta _{km1}}} \right), \cdots ,{\mathbf{a}}\left( {{\theta _{km{N_m}}}} \right)} \right]\in {\mathbb C}^{L\times N_m}$, with ${\left[ {{\bf{a}}\left( \theta  \right)} \right]_{l'}} = {e^{j\theta \left(l'-1\right)}} = {e^{ - j2\pi \frac{\tau }{{LT}}\left( {l' - 1} \right)}}$ and ${{\mathbf{z}}_{km}} = {\left[ {{z_{km1}}, \cdots ,{z_{km{N_m}}}} \right]^T}$. To start the inference, the probability $p\left( {{\mathcal{H}_{km}},\{ {\theta _{kmn}}\} ,\{ {z_{kmn}}\} } \right)$ could be factorized as
\begin{equation}
	\begin{aligned}
		&p\left( {{\mathcal{H}_{km}},\{ {\theta _{kmn}}\} ,\{ {z_{kmn}}\} } \right)=\\
		& \mathcal{C}\mathcal{N}\left( {{\mathcal{H}_{km}},{{\mathbf{A}}_{km}}\left( {{{\boldsymbol{\theta }}_{km}}} \right){\mathbf{z}},\sigma _{km}^2{{\mathbf{I}}_L}} \right)\prod\limits_{n = 1}^{{N_m}} {p\left( {{\theta _{kmn}}} \right)p\left( {{z_{kmn}}} \right)}.
	\end{aligned}
\end{equation}

By inserting auxiliary nodes ${{\mathbf{s}}_{kmn}} = {\mathbf{a}}\left( {{\theta _{kmn}}} \right){z_{kmn}}$, the factorization is transformed into Eq.(\ref{factorization}) and the factor graph is depicted in Fig. \ref{fig:FactorGraph}.

In the left-hand segment (main branch) of Fig. \ref{fig:FactorGraph}, 
the Gaussian sub-problem is addressed using BP, while the non-linear inference associated with the individual 
sub-blocks of ${\bf s}_{kmn}$ is carried out by invoking variational inference \cite{VALSE,VMP,PAML}. 
It is worth noting that this arrangement constitutes a hybrid inference framework, 
wherein the inference processes within each sub-block are executed independently, 
yet a limited number of probabilistic messages are exchanged between the sub-blocks 
and the main branch through the node $f_{kmn}$, thereby maintaining statistical coupling across the system. 
Furthermore, the random variables $\{\theta_{kmn}\}$ and $\{z_{kmn}\}$ are assumed 
to remain mutually independent throughout the inference procedure. To mitigate the 
computational burden imposed by high-dimensional covariance updates, a scalar-type 
approximation is employed, namely, the covariance matrices are constrained to be 
scaled identity matrices, i.e., proportional to ${\bf I}$.

\begin{figure*}[t]
	\begin{equation}\label{factorization}
		p\left( {{{\cal H}_{km}},\{ {\theta _{kmn}}\} ,\{ {z_{kmn}}\} } \right) = \underbrace {{\cal C}{\cal N}\left( {{{\cal H}_{km}},\sum\limits_{n = 1}^{{N_m}} {{{\bf{s}}_{kmn}}} ,\sigma _{km}^2{{\bf{I}}_L}} \right)}_{{\text{Approximated Belief Propagation Inference}}}\underbrace {\prod\limits_{n = 1}^{{N_m}} {\delta \left( {{{\bf{s}}_{kmn}} - {\bf{a}}\left( {{\theta _{kmn}}} \right){z_{kmn}}} \right)p\left( {{\theta _{kmn}}} \right)p\left( {{z_{kmn}}} \right)} }_{{\text{Approximated Variational Inference}}},
	\end{equation}
\end{figure*}

\begin{figure*}[t]
	\begin{equation}\label{sumforward}
		{\mu _{{{\cal H}_{km}} \to {{\bf{s}}_{kmn}}}}\left( {{{\bf{s}}_{kmn}}} \right) \propto \prod\limits_{n' \ne n}^{{N_m}} {{\mu _{{{\cal H}_{km}} \leftarrow {{\bf{s}}_{kmn'}}}}\left( {{{\bf{s}}_{kmn'}}} \right)}
		{\cal C}{\cal N}\left( {{{\cal H}_{km}};\sum\limits_{n' \ne n}^{{N_m}} {{{\bf{s}}_{kmn'}} + {{\bf{s}}_{kmn}}} ,\sigma _{km}^2{{\bf{I}}_L}} \right)
	\end{equation}
	\noindent\rule{\textwidth}{0.4pt}
\end{figure*}

\begin{figure}[!t]
	\centering
	\includegraphics[width=0.8\linewidth]{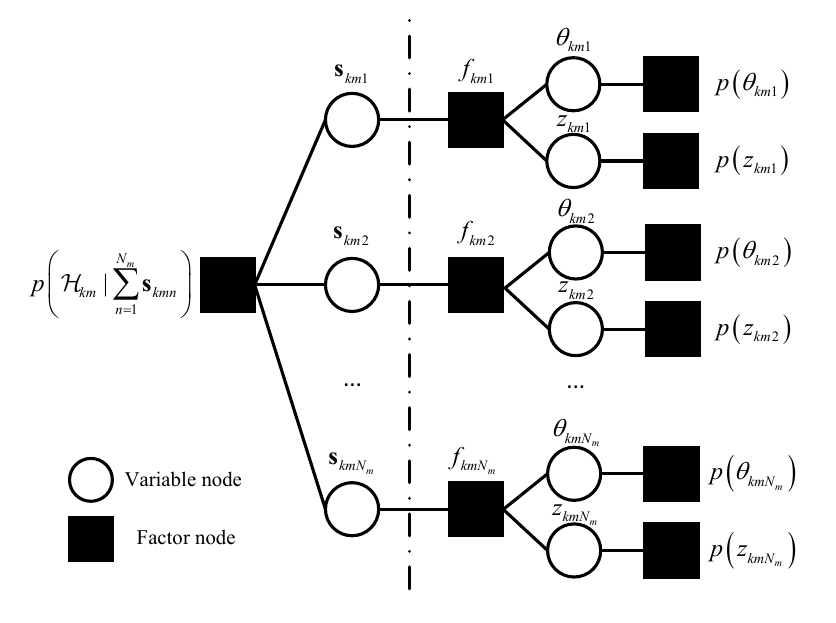}
	\centering
	\caption{The factor graph of \eqref{factorization}; on the left of the dashed line, we adopt the belief propagation, while in the right side of the dashed line, the variational inference is adopted.}
	\label{fig:FactorGraph}
\end{figure}

In the absence of explicit \emph{a priori} knowledge concerning the parameters 
$\theta_{kmn}$ and $z_{kmn}$, suitable postulated prior distributions are nevertheless 
required in order to facilitate the ensuing Bayesian inference. Accordingly, the prior 
PDF of the phase-related variable $\theta_{kmn}$ is modeled 
by a Von~Mises distribution $p(\theta_{kmn}) = \mathcal{VM}(\theta_{kmn}; 0, \zeta_{kmn})$, where $\zeta_{kmn}$ denotes the corresponding concentration parameter. In parallel, 
the complex-valued coefficient $z_{kmn}$ is assumed to obey a circularly symmetric 
complex Gaussian prior of the form $p(z_{kmn}) = \mathcal{CN}(z_{kmn}; 0, \xi_{kmn}^2)$,
where $\xi_{kmn}^2$ characterizes its variance. Since these priors are introduced solely 
for mathematical tractability and do not convey genuine prior information, the variance 
parameter $\xi_{kmn}$ is deliberately chosen to be sufficiently large, whereas the 
concentration factor $\zeta_{kmn}$ (or equivalently the precision $\kappa_{kmn}$) is set 
to a relatively small value. This ensures that the priors remain non-informative, thereby 
allowing the inference process to be predominantly governed by the observed data.

To start, the forward message ${\mu _{{{\cal H}_{km}} \to {{\bf{s}}_{kmn}}}}\left( {{{\bf{s}}_{kmn}}} \right)$ is Eq. (\ref{sumforward}).
Given that ${\mu _{{{\cal H}_{km}} \leftarrow {{\bf{s}}_{kmn'}}}}\left( {{{\bf{s}}_{kmn'}}} \right) = {\cal C}{\cal N}\left( {{{\bf{s}}_{kmn'}};{{{\bf{\hat s}}}_{{{\cal H}_{km}} \leftarrow {{\bf{s}}_{kmn'}}}},{\sigma _{{{\cal H}_{km}} \leftarrow {{\bf{s}}_{kmn'}}}^2}{\bf I}_L} \right)$, $n'\neq n$ are independent Gaussian messages, then we have ${\mu _{{{\cal H}_{km}} \to {{\bf{s}}_{kmn}}}}\left( {{{\bf{s}}_{kmn}}} \right) = {\cal C}{\cal N}\left( {{{\bf{s}}_{kmn'}};{{{\bf{\hat s}}}_{{{\cal H}_{km}} \to {{\bf{s}}_{kmn}}}},{\sigma _{{{\cal H}_{km}} \to {{\bf{s}}_{kmn}}}^2}{\bf I}_L} \right)$, with
\begin{equation}
	\sigma _{{{\cal H}_{km}} \to {{\bf{s}}_{kmn}}}^2 = \sigma _{km}^2 + \sum\limits_{n' \ne n}^{{N_m}} {\sigma _{{{\cal H}_{km}} \leftarrow {{\bf{s}}_{kmn'}}}^2} 
\end{equation}
and
\begin{equation}
	{{{\bf{\hat s}}}_{{{\cal H}_{km}} \to {{\bf{s}}_{kmn}}}} = {{\cal H}_{km}} - \sum\limits_{n' \ne n}^{{N_m}} {{{{\bf{\hat s}}}_{{{\cal H}_{km}} \leftarrow {{\bf{s}}_{kmn'}}}}}.
\end{equation}

In the block-wise variational inference stage, the constraint function ${f_{kmn}}$ 
is derived by fusing the incoming message of ${{\bf s}_{kmn}}$ with its structural 
constraint. Specifically, we have
\begin{equation}
	\label{eq:fkmn_def}
	\begin{aligned}
		f_{kmn} 
		&= \int_{{\bf s}_{kmn}} 
		{\mu_{{{\cal H}_{km}} \to {\bf s}_{kmn}}\!\left( {\bf s}_{kmn} \right)
			\delta \!\left( {\bf s}_{kmn} - {\bf a}\!\left( \theta_{kmn} \right) z_{kmn} \right)} \\
		&= {\cal CN}\!\left( 
		{\bf a}\!\left( \theta_{kmn} \right) z_{kmn};
		{\bf{\hat s}}_{{{\cal H}_{km}} \to {\bf s}_{kmn}},
		\sigma_{{{\cal H}_{km}} \to {\bf s}_{kmn}}^2 {\bf I}_L
		\right),
	\end{aligned}
\end{equation}
where ${\bf{\hat s}}_{{{\cal H}_{km}} \to {\bf s}_{kmn}}$ and 
$\sigma_{{{\cal H}_{km}} \to {\bf s}_{kmn}}^2$ denote the mean and variance 
of the Gaussian message passed from module ${{\cal H}_{km}}$ to 
${\bf s}_{kmn}$, respectively. 

Consequently, the outgoing message from the constraint node to the phase variable 
$\theta_{kmn}$, denoted as ${\mu_{{f_{kmn}} \to \theta_{kmn}}}\!\left( \theta_{kmn} \right)$, 
can be expressed as
\begin{equation}
	\label{eq:g_def}
	{\mu_{{f_{kmn}} \to \theta_{kmn}}}\!\left( \theta_{kmn} \right) 
	\propto 
	\exp\!\left(
	g_{{f_{kmn}} \to \theta_{kmn}}\!\left( \theta_{kmn} \right)
	\right),
\end{equation}
where
\begin{equation}
	g_{{f_{kmn}} \to \theta_{kmn}}\!\left( \theta_{kmn} \right)
	= 
	\int_{z_{kmn}} 
	\ln \! \left[ f_{kmn} \,
	{\mu_{{f_{kmn}}}}\!\left( z_{kmn} \right) \right] \mathrm{d}z_{kmn}.
\end{equation}

Assuming that the complex-valued message of $z_{kmn}$ follows 
${\mu_{{z_{kmn}}}}\!\left( z_{kmn} \right) 
= {\cal CN}\!\left( z_{kmn}; {\hat z}_{{f_{kmn}}}, \sigma_{{f_{kmn}}}^2 \right)$,
we ultimately arrive at Eq. (\ref{thetap}).
\begin{figure*}[tb]
	\begin{equation}\label{thetap}
		{\mu_{{f_{kmn}} \to \theta_{kmn}}}\!\left( \tau_{kmn} \right)
		\propto 
		\exp \!\left(
		2 \sigma_{{{\cal H}_{km}} \to {\bf s}_{kmn}}^{-2} 
		\, \Re \! \left\{ 
		{\bf{\hat s}}_{{{\cal H}_{km}} \to {\bf s}_{kmn}}^{\!H}
		{\bf a}\!\left( \theta_{kmn} \right)
		{\hat z}_{{f_{kmn}} \leftarrow z_{kmn}}
		\right\}
		\right)
	\end{equation}
\end{figure*}

The resultant expression represents the product of multiple folded 
Von~Mises (VM) distributions\cite{PAML}, which can be well approximated 
by a single equivalent VM distribution through a second-order Taylor 
expansion of the modified phase in \eqref{thetap}, as \[{q_{{\text{MVM}}}} = {\Re}\left( {2\sigma _{{H_{km}} \to {{\mathbf{s}}_{kmn}}}^{ - 2} {{\mathbf{\hat s}}_{{H_{km}} \to {{\mathbf{s}}_{kmn}}}^{H}\left( {{\mathbf{a}}\left( \theta  \right) - {{\mathbf{1}}_{L \times 1}}/L} \right){{\hat z}_{{f_{kmn}} \leftarrow {z_{kmn}}}}} } \right)\] around the aligned 
phase ${{\bar \theta}_{{f_{kmn}} \to \theta_{kmn}}}$. Hence, the message 
${\mu_{{f_{kmn}} \to \theta_{kmn}}}\!\left( \theta_{kmn} \right)$ 
can be approximated as
\begin{equation}
	{\mu_{{f_{kmn}} \to \theta_{kmn}}}\!\left( \theta_{kmn} \right)
	= 
	{\cal VM}\!\left( 
	\theta_{kmn}; 
	{\hat \theta}_{{f_{kmn}} \to \theta_{kmn}}, 
	{\kappa}_{{f_{kmn}} \to \theta_{kmn}}
	\right),
\end{equation}
where ${\hat \theta}_{{f_{kmn}} \to \theta_{kmn}} =
{\bar \theta}_{{f_{kmn}} \to \theta_{kmn}} 
- 
\frac{q_{\mathrm{MVM}}'\!\left( 
	{\bar \theta}_{{f_{kmn}} \to \theta_{kmn}} \right)}
{q_{\mathrm{MVM}}''\!\left( 
	{\bar \theta}_{{f_{kmn}} \to \theta_{kmn}} \right)} 
$ and ${\kappa}_{{f_{kmn}} \to \theta_{kmn}}
= 
f_I^{-1}\!\left( 
\exp \!\left[
-\frac{1}{
	2 q_{\mathrm{MVM}}''\!\left( 
	{\bar \theta}_{{f_{kmn}} \to \theta_{kmn}} \right)}
\right]
\right)$ depicts the quadrature approximation around ${{\bar \theta}_{{f_{kmn}} \to \theta_{kmn}}}$
and $f_I(\cdot) = I_1(\cdot)/I_0(\cdot)$, with $I_0(\cdot)$ and $I_1(\cdot)$ denoting 
the modified Bessel functions of the first kind. The aligned phase 
${\bar \theta}_{{f_{kmn}} \to \theta_{kmn}}$ can be efficiently estimated by 
maximizing ${\bf{\hat s}}_{{{\cal H}_{km}} \to {\bf s}_{kmn}}^{H}
{\bf a}\!\left( \theta \right)$ via the modified greedy unwrap algorithm in \cite{VALSE}.

Afterwards, the backward message from variable node ${{\theta _{kmn}}}$ is ${\mu _{ {\theta _{kmn}}}}\left( {{\theta _{kmn}}} \right) = {\cal V}{\cal M}\left( {{\theta _{kmn}};{{\hat \theta }_{{f_{kmn}}}},{\kappa _{{\theta _{kmn}}}}} \right)$ with ${\kappa _{ {\theta _{kmn}}}} = \left| {{\eta _{kmn}}} \right|$ and ${{\hat \theta }_{ {\theta _{kmn}}}} = \mathrm{angle}\left( {{\eta _{kmn}}} \right)$ where ${\eta _{kmn}} = {\kappa _{{f_{kmn}} \to {\theta _{kmn}}}}\exp\left( {j{{\hat \theta }_{{f_{kmn}} \to {\theta _{kmn}}}}} \right) + {\zeta _{{{kmn}}}}$.

Then, for the message from ${{f_{kmn}}}$ to ${{z _{kmn}}}$, we have 
\begin{equation}
	\begin{aligned}
		{\mu _{{f_{kmn}} \to {z_{kmn}}}}\left( {{z_{kmn}}} \right) \propto \exp \left( {g_{{f_{kmn}} \to {z_{kmn}}}}\left( {{z_{kmn}}} \right) \right),
	\end{aligned}
\end{equation}
with 
\begin{equation}
	\begin{aligned}
		&{g_{{f_{kmn}} \to {z_{kmn}}}}\left( {{z_{kmn}}} \right)
		=\int\limits_{{\theta _{kmn}}} {\ln {f_{kmn}}{\mu _{{\theta _{kmn}}}}\left( {{\theta _{kmn}}} \right)} \\
		&= -\left( L{{ {{z_{kmn}^H}}}}{z_{kmn}}+  2{\mathop \Re\nolimits} \left( {{\bf{\hat s}}_{{{\cal H}_{km}} \to {{\bf{s}}_{kmn}}}^H{\bf{\hat a}}\left( {{\theta _{kmn}}} \right){z_{kmn}}} \right) \right)\\
		&\times \sigma _{{{\cal H}_{km}} \to {{\bf{s}}_{kmn}}}^{ - 2}+\mathrm{const},
	\end{aligned}
\end{equation}
where ${\left[ {{\bf{\hat a}}\left( \theta  \right)} \right]_l} = {\left[ {{\mathbb E}\left( {{\bf{a}}\left( \theta  \right)} \right)} \right]_l} = \frac{{{I_{l - 1}}\left( {{\kappa _\theta }} \right)}}{{{I_0}\left( {{\kappa _\theta }} \right)}}{\bf{a}}\left( {\hat \theta } \right)$ for $\theta \sim {\cal V}{\cal M}\left( {\theta ;\hat \theta ,{\kappa _\theta }} \right)$. Note that ${g_{{f_{kmn}} \to {z_{kmn}}}}\left( {{z_{kmn}}} \right)$ is a typical complex quadratic function, hence ${\mu _{{f_{kmn}} \to {z_{kmn}}}}\left( {{z_{kmn}}} \right) \propto {\cal C}{\cal N}\left( {{z_{kmn}};{{\hat z}_{{f_{kmn}} \to {z_{kmn}}}},\sigma _{{f_{kmn}} \to {z_{kmn}}}^2} \right)$, with
\begin{equation}
	\sigma _{{f_{kmn}} \to {z_{kmn}}}^2 = \sigma _{{{\cal H}_{km}} \to {{\bf{s}}_{kmn}}}^2/L,
\end{equation}
and 
\begin{equation}
	{{\hat z}_{{f_{kmn}} \to {z_{kmn}}}} = {{{\bf{\hat a}}}^H}\left( {{\theta _{kmn}}} \right){{{\bf{\hat s}}}_{{{\cal H}_{km}} \to {{\bf{s}}_{kmn}}}}/L.
\end{equation}

Therefore, the parameters of the backward message ${\mu _{ {z_{kmn}}}}\left( {{z_{kmn}}} \right)$ could be derived by ${{\hat z}_{{z_{kmn}}}} = \sigma _{{z_{kmn}}}^2/\sigma _{{f_{kmn}} \to {z_{kmn}}}^2{{\hat z}_{{f_{kmn}} \to {z_{kmn}}}}$ and $\sigma _{{z_{kmn}}}^2 = {\left( {\sigma _{{{\cal H}_{km}} \to {{\bf{s}}_{kmn}}}^{ - 2}L + \xi _{kmn}^{ - 2}} \right)^{ - 1}}$.

The approximated Gaussian estimate of ${\bf s}_{kmn}$ is then generated as ${{\bf{s}}_{kmn}}\sim{\cal C}{\cal N}\left( {{{\bf{s}}_{kmn}};{{{\bf{\hat s}}}_{kmn}},\sigma _{{{\bf{s}}_{kmn}}}^2{{\bf{I}}_L}} \right)$ with ${{{\bf{\hat s}}}_{kmn}} = {\mathbb E}\left( {{z_{kmn}}{\bf{a}}\left( {{\theta _{kmn}}} \right)} \right) = {{\hat z}_{kmn}}{\bf{\hat a}}\left( {{\theta _{kmn}}} \right)$ and \[\sigma _{{{\bf{s}}_{kmn}}}^2 \approx \frac{1}{L}{\rm{tr}}\left[ {{\mathbb E}\left( {{z_{kmn}}{\bf{a}}\left( {{\theta _{kmn}}} \right){{\left( {{z_{kmn}}{\bf{a}}\left( {{\theta _{kmn}}} \right)} \right)}^H}} \right) - {{{\bf{\hat s}}}_{kmn}}{\bf{\hat s}}_{kmn}^H} \right] = \frac{1}{L}\sigma _{{z_{kmn}}}^2.\] By deducting the incoming message, the parameters of ${\mu _{{{\cal H}_{km}} \leftarrow {{\bf{s}}_{kmn}}}}\left( {{{\bf{s}}_{kmn}}} \right)$ could be derived as
${{{\bf{\hat s}}}_{{{\cal H}_{km}} \leftarrow {{\bf{s}}_{kmn}}}} = \sigma _{{{\cal H}_{km}} \leftarrow {{\bf{s}}_{kmn}}}^2{{\bf{I}}_L}\left( {\sigma _{{{\bf{s}}_{kmn}}}^{ - 2}{{{\bf{\hat s}}}_{kmn}} - \sigma _{{{\cal H}_{km}} \to {{\bf{s}}_{kmn}}}^{ - 2}{{{\bf{\hat s}}}_{{{\cal H}_{km}} \to {{\bf{s}}_{kmn}}}}} \right)$
and $\sigma _{{{\cal H}_{km}} \leftarrow {{\bf{s}}_{kmn}}}^2 = {\left( {\sigma _{{{\bf{s}}_{kmn}}}^{ - 2} - \sigma _{{{\cal H}_{km}} \to {{\bf{s}}_{kmn}}}^{ - 2}} \right)^{ - 1}}$.

To further mitigate the detrimental impact of spectral artifacts, i.e., the spurious 
spectral lines that are closely spaced within a predefined threshold ${\epsilon_\theta}$, 
a successive spectral line fusion procedure is invoked. Specifically, during each iteration, i.e.,
whenever two estimated spectral components satisfy 
$\left\| {\hat \theta_{kmn_1} - \hat \theta_{kmn_2}} \right\| < {\epsilon_\theta}$, 
they are regarded as representing the same physical spectral line. A refined estimate is 
then obtained as
\begin{equation}
	\label{eq:theta_fusion}
	\hat \theta_{kmn_1}^{\mathrm{new}}
	= 
	\mathrm{angle}\!\left(
	\kappa_{kmn_1} e^{j \hat \theta_{kmn_1}}
	+ 
	\kappa_{kmn_2} e^{j \hat \theta_{kmn_2}}
	\right),
\end{equation}
and the corresponding concentration parameter is updated according to
\begin{equation}
	\label{eq:kappa_fusion}
	\kappa_{kmn_1}^{\mathrm{new}}
	=
	\left\|
	\kappa_{kmn_1} e^{j \hat \theta_{kmn_1}}
	+
	\kappa_{kmn_2} e^{j \hat \theta_{kmn_2}}
	\right\|.
\end{equation}
Here, $\hat \theta_{kmn_1}^{\mathrm{new}}$ and 
$\kappa_{kmn_1}^{\mathrm{new}}$ represent the updated mean and 
concentration parameters, respectively, replacing their previous counterparts.
Similarly, the gain parameters $z_{kmn_1}$ and $z_{kmn_2}$ are merged 
in a statistically consistent manner as
$
\left( \sigma_{z_{kmn_1}}^{\mathrm{new}} \right)^2
=
\left(
\sigma_{z_{kmn_1}}^{-2}
+
\sigma_{z_{kmn_2}}^{-2}
\right)^{-1},$
and
\[\hat z_{kmn_1}^{\mathrm{new}}
=
\left( \sigma_{z_{kmn_1}}^{\mathrm{new}} \right)^2
\left(
\sigma_{z_{kmn_1}}^{-2} \hat z_{kmn_1}
+
\sigma_{z_{kmn_2}}^{-2} \hat z_{kmn_2}
\right)\], 
ensuring that both the mean and the variance of the fused coefficient are properly weighted.

During the subsequent iteration, the parameters of the newly formed messages are 
re-initialized as
\begin{equation}
	\label{eq:sigma_renew}
	\sigma_{{\mathcal H}_{km} \to {\bf s}_{kmn_{\mathrm{new}}}}^2
=
\sigma_{km}^2
+
\sum_{n'=1}^{n_{\mathrm{new}}-1}
\sigma_{{\mathcal H}_{km} \leftarrow {\bf s}_{kmn'}}^2,
\end{equation}
\begin{equation}
\label{eq:s_renew}
{\bf{\hat s}}_{{\mathcal H}_{km} \to {\bf s}_{km_{\mathrm{new}}}}
=
{\mathcal H}_{km}
-
\sum_{n'=1}^{n_{\mathrm{new}}-1}
{\bf{\hat s}}_{{\mathcal H}_{km} \leftarrow {\bf s}_{kmn'}},
\end{equation}
for $n_{\mathrm{new}} = N_{\mathrm{valid}} + 1, \ldots, N_m$, 
where $N_{\mathrm{valid}}$ denotes the number of valid messages 
retained from the previous iteration. This successive fusion strategy 
effectively suppresses redundant spectral components while maintaining 
the integrity of the valid frequency estimates. 

From the similarity between the VM distribution in $[-\pi,\pi]$ and the Gaussian distribution, the final estimate of $\tau_{kmn}$, $n-1,\cdots, N_m$ could be approximated as ${\tau _{kmn}}\sim\mathcal{N}\left( {{\tau _{kmn}};{{\hat \tau }_{kmn}},\sigma _{{\tau _{kmn}}}^2} \right)$, with ${{\hat \tau }_{kmn}} =  - \frac{{LT}}{{2\pi }}{{\hat \theta }_{kmn}}$ and $\sigma _{{\tau _{kmn}}}^2 = \frac{{{L^2}{T^2}}}{{4{\pi ^2}}}{\kappa _{kmn}}$.

The full procedure is depicted in Algorithm \ref{BP_VMPalg}.

\begin{figure*}[t]
	\begin{equation}
		\label{locmessage}
		\begin{aligned}
			{\mu_{\boldsymbol{\psi}_k}}(x_k,y_k)
			&\propto
			\int_{\{\tau_{kmn}\}}
			\delta\!\Big(\tau_{kmn}-\tau_{mn}^{i}-\tau_{kmn}^{o}(x_k,y_k)\Big)
			\prod_{m=1}^{M}\prod_{n=1}^{N_m}
			\mathcal{N}\!\left(\tau_{kmn};\hat{\tau}_{kmn},\sigma_{\tau_{kmn}}^{2}\right)\\
			&=
			\prod_{m=1}^{M}\prod_{n=1}^{N_m}
			\mathcal{N}\!\left(\tau_{kmn}(x_k,y_k);\hat{\tau}_{kmn},\sigma_{\tau_{kmn}}^{2}\right).
		\end{aligned}
	\end{equation}
	\noindent\rule{\textwidth}{0.4pt}
\end{figure*}

\begin{algorithm}[t]
	\caption{Message Passing Based Delay Extraction for ${\mathcal{H}_{km}}$}
	\begin{algorithmic}[1]
		\Require ${\mathcal{H}_{km}}$, $N_m$, $\epsilon_{\theta}$
		\Ensure  ${\mathbf{\hat z}}_{km} \in \mathbb{C}^{N_m}$, ${\mathbf{\hat \tau}}_{kmn} \in \mathbb{R}^{N_m}$
		\State Initialize $\sigma _{{{\cal H}_{km}} \leftarrow {{\bf{s}}_{kmn}}}^2$ and  ${{{\bf{\hat s}}}_{{{\cal H}_{km}} \leftarrow {{\bf{s}}_{kmn}}}}$ for $n=1,\cdots,N$.
		\For{$n_{iter} = 1,\cdots,N_{it}^{in}$}
		\For{$n = 1,\cdots,N_m$ }
		\State Calculate ${\sigma _{{{\cal H}_{km}} \to {{\bf{s}}_{kmn}}}^{2}}$ and ${{{{\bf{\hat s}}}_{{{\cal H}_{km}} \to {{\bf{s}}_{kmn}}}}}$;
		\State Calculate ${{\hat \theta }_{{f_{kmn}} \to {\theta _{kmn}}}}$ and ${\kappa _{{f_{kmn}} \to {\theta _{kmn}}}}$;
		\State Calculate ${{\hat \theta }_{kmn}}$ and ${\kappa_{{\theta_{kmn}}}}$;
		\State Calculate ${{\hat z }_{{f_{kmn}} \to {z _{kmn}}}}$ and ${\sigma _{{f_{kmn}} \to {z _{kmn}}}^2}$;
		\State Calculate ${{\hat z }_{{z _{kmn}}}}$ and ${\sigma _{{z _{kmn}}}^2}$;
		\State Calculate ${{{{\bf{\hat s}}}_{kmn}}}$ and ${\sigma _{{{\bf{s}}_{kmn}}}^2}$;
		\State Calculate ${{{\bf{\hat s}}}_{{{\cal H}_{km}} \leftarrow {{\bf{s}}_{kmn}}}}$ and $\sigma _{{{\cal H}_{km}} \leftarrow {{\bf{s}}_{kmn}}}^2$;
		\EndFor
		\State Merge the artifacts if $\left\| {{\hat \theta _{km{n_2}}} - {\hat \theta _{km{n_1}}}} \right\| < {\epsilon_\theta }$, for $n_2 \neq n_1$;
		\EndFor
		\State Acquire $\hat\tau_{kmn}$ and $\sigma _{{\tau _{kmn}}}^2$ from ${{\hat \theta }_{kmn}}$ and ${\kappa_{{\theta_{kmn}}}}$ for $n=1,\cdots,N_m$;
		\State \Return $\hat\tau_{kmn}$, $\sigma _{{\tau _{kmn}}}^2$, ${{\hat z }_{{z _{kmn}}}}$ and ${\sigma _{{z _{kmn}}}^2}$.
	\end{algorithmic}
	\label{BP_VMPalg}
\end{algorithm}

\subsection{Localization Using Delays}

Upon obtaining the delay estimates $\{\tau_{kmn}\}$ for the $k$-th user,
with $m=1,\ldots,M$ and $n=1,\ldots,N_m$, via either the proposed
dictionary-based estimator or the message-passing-based scheme, we proceed to
refine the position vector $\boldsymbol{\psi}_k=[x_k,\,y_k,\,0]^T$.
The path delay is decomposed as
\begin{equation}
	\tau_{kmn}(x_k,y_k) = \tau_{mn}^{i} + \tau_{kmn}^{o}(x_k,y_k),
\end{equation}
where $\tau_{mn}^{i}$ denotes the intrinsic delay, while
the geometric delay $\tau_{kmn}^{o}(x_k,y_k)$ obeys
$\tau_{kmn}^{o}(x_k,y_k)
=
\frac{\left\|\boldsymbol{\psi}_k-\boldsymbol{\psi}_{mn}^{\mathrm{anc}}\right\|}{c}
=
\frac{1}{c}
\sqrt{(x_k-x_{mn}^{\mathrm{anc}})^2+(y_k-y_{mn}^{\mathrm{anc}})^2+(z_{mn}^{\mathrm{anc}})^2}.$

Since the observed delays $\{\tau_{kmn}\}$ are inherently unordered (i.e., not
a priori associated with the anchors/PAs $\{\boldsymbol{\psi}_{mn}^{\mathrm{anc}}\}$),
a delay-anchor matching step is required. To this end, for each sink $m$ we form
the reference delays $\tau_{kmn}'=\tau_{kmn}(x_0,y_0)$ at an initial point
$(x_0,y_0)$ and solve the assignment with the Hungarian algorithm using the
$N_m\times N_m$ cost matrix ${\bf C}_m$ whose $(n_1,n_2)$-th entry is $c^{m}_{n_1,n_2}
=
\log\!\left(
1+\frac{\left|\tau_{kmn_1}-\tau_{kmn_2}'\right|^2}{\delta}
\right),$
where $\delta$ is a scaling factor controlling the sensitivity of the matcher.

Conditioned on the ordered Gaussian delay posteriors delivered by Algorithm~\ref{BP_VMPalg},
the marginal of $\boldsymbol{\psi}_k$ follows the sum–product rule as \eqref{locmessage}.

To obtain a tractable refinement, we linearize $\tau_{kmn}(x,y)$ at
$(x_0,y_0)$ as $\tau_{kmn}(x,y)\approx g_{kmn}(x,y)
=
c_{kmn}(x_0,y_0)
+
a_{kmn}(x_0,y_0)(x)
+
b_{kmn}(x_0,y_0)(y)$ with spatial gradients evaluated at $(x_0,y_0)$ defined as
\begin{align}
	a_{kmn} &\triangleq
	\left.\frac{\partial \tau_{kmn}(x,y)}{\partial x}\right|_{(x_0,y_0)}
	=
	\frac{x_k-x_{mn}^{\mathrm{anc}}}{c\left\|\boldsymbol{\psi}_k-\boldsymbol{\psi}_{mn}^{\mathrm{anc}}\right\|},\\
	b_{kmn} &\triangleq
	\left.\frac{\partial \tau_{kmn}(x,y)}{\partial y}\right|_{(x_0,y_0)}
	=
	\frac{y_k-y_{mn}^{\mathrm{anc}}}{c\left\|\boldsymbol{\psi}_k-\boldsymbol{\psi}_{mn}^{\mathrm{anc}}\right\|},
\end{align}
with the offset
\begin{equation}
	c_{kmn}\triangleq \tau_{kmn}(x_0,y_0)-a_{kmn}x_0-b_{kmn}y_0.
\end{equation}
Note that ${\mu_{\boldsymbol{\psi}_k}}(x_k,y_k)$ with linearized $\tau_{kmn}$
 is jointly Gaussian, as ${\mu_{\boldsymbol{\psi}_k}}(x_k,y_k)\;\propto\;
\mathcal{N}\!\Big(\,[x_k,\,y_k]^T;\,{\bf p}_{\boldsymbol{\psi}_k},\,{\bf\Sigma}_{\boldsymbol{\psi}_k}\Big)$ 
with 
\begin{equation}
	\label{eq:locsigma}
	{\bf\Sigma}_{\boldsymbol{\psi}_k}
	=
	\sum_{m=1}^{M}\sum_{n=1}^{N_m}{\bf\Sigma}_{\boldsymbol{\psi}_kmn},
\end{equation}
\begin{equation}\label{eq:locp}
	{\bf p}_{\boldsymbol{\psi}_k}
	=
	{\bf\Sigma}_{\boldsymbol{\psi}_k}^{-1}
	\sum_{m=1}^{M}\sum_{n=1}^{N_m}{\bf p}_{\boldsymbol{\psi}_kmn},
\end{equation}
where
\begin{equation}
	\label{eq:locsigmasub}
	{\bf\Sigma}_{\boldsymbol{\psi}_kmn}
	=
	\sigma_{\tau_{kmn}}^{-2}
	\begin{bmatrix}
		|a_{kmn}|^{2} & a_{kmn}b_{kmn}\\
		a_{kmn}b_{kmn} & |b_{kmn}|^{2}
	\end{bmatrix},
\end{equation}
\begin{equation}\label{eq:locpsub}
	{\bf p}_{\boldsymbol{\psi}_kmn}
	=
	\big(\hat\tau_{kmn}-c_{kmn}\big)\,\sigma_{\tau_{kmn}}^{-2}
	\begin{bmatrix}
		a_{kmn}\\[2pt]
		b_{kmn}
	\end{bmatrix}.
\end{equation}
With a sufficiently rich set of measurements, the estimates $(\hat x_k,\hat y_k)$
could be obtained approximately in a closed form from~\eqref{eq:locsigma}-\eqref{eq:locpsub}. For a non-Bayesian delay extractor, let all candidates be assumed to share the same
variance, the variance term in~\eqref{eq:locsigmasub} effectively collapses and a
point estimate is produced.

\begin{figure}[!t]
	\centering
	\includegraphics[width=0.8\linewidth]{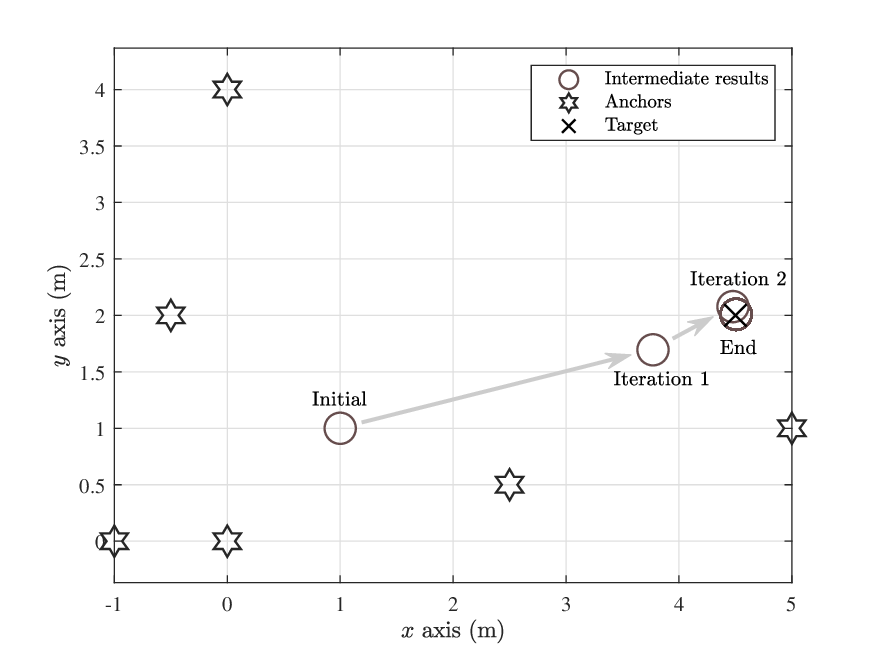}
	\centering
	\caption{A graphical illustration of the iterative delay based localization, where the noisy observation of the delays are generated by directly adding AWGN with variance $10^{-10}$, and the initial point of iteration is $\left(1,1,0\right)$.}
	\label{fig:Lociter}
\end{figure}

Since the linearization $g_{kmn}(x,y)$ is a first-order surrogate and the
delay–anchor association hinges on $(x_0,y_0)$, a Newton
refinement is invoked:

\begin{enumerate}
	\item Using $\tau_{kmn}(x_0,y_0)$, perform delay–anchor matching; linearize
	$\tau_{kmn}(x,y)$ at $(x_0,y_0)$ and compute
	$(\hat x_k,\hat y_k)$ via~\eqref{eq:locsigma}–\eqref{eq:locpsub}.
	\item Re-evaluate $\tau_{kmn}(x,y)$ at $(\hat x_k,\hat y_k)$, re-run matching,
	and update $(\hat x_k,\hat y_k)$ accordingly.
	\item Iterate Step~2 until the change in $(\hat x_k,\hat y_k)$ falls below a given threshold.
\end{enumerate}
Here, a graphic illustration of the iterative refinement procedure is depicted in Fig. \ref{fig:Lociter}.

The estimation of $\bf h$, ${{{{\mathbf{\hat h}}}^o}}$ and ${\sigma _h^o}$ could then be obtained via the delta method. By concatenating the Gaussian estimates of ${\bf z}_k$ and ${\bm \psi}_k$, the real variable set ${\bm{\zeta}} = \left[ \Re \left( {{{\mathbf{z}}^T}} \right),\Im \left( {{{\mathbf{z}}^T}} \right),\{x_k\}_{1}^{K},\{y_k\}_{1}^{K} \right]^T$ is jointly Gaussian distributed with mean ${{\bm{\hat \zeta}}}$ and covariance matrix ${{\mathbf{\Sigma }}_{\bm{\zeta}}}$. Then the mean ${{{\mathbf{\hat h}}}^{\text {out}}}$ and scalar variance ${\left( {\sigma _h^{\text {out}}} \right)^2}$ could be approximated as ${{{\mathbf{\hat h}}}^{\text {out}}} = {\mathbf{g}_{\bf h}}\left( {{\bm{\hat \zeta}}} \right)$ and ${\left( {\sigma _h^{\text {out}}} \right)^2} = \frac{1}{{KML}}\text{tr}\left[ {\frac{{\partial {\mathbf{g}_{\bf h}}}}{{\partial {\bm{\zeta}}}}{{\mathbf{\Sigma }}_{{\bm{\zeta}}}}{{\left( {\frac{{\partial {\mathbf{g}_{\bf h}}}}{{\partial {\bm{\zeta}}}}} \right)}^T}} \right]$, respectively, where ${\mathbf{g}_{\bf h}}:{\mathbb R}^{2K+2KN_{all}}\mapsto{\mathbb R}^{2KML}$ maps the parameters to the channel. For non-Bayesian methods (e.g., Algorithm~\ref{OMPalg}), a point estimate
$\hat{\bf h}^{\,o}$ is obtained directly, while the variance can be approximated
from the whole estimator's divergence, thereby yielding a consistent uncertainty proxy.

\section{Theoretical Analysis}

\subsection{Cramér–Rao Lower Bound}

The achievable estimation accuracy of the proposed framework can be
benchmarked using the CRLB, which provides a
fundamental lower limit on the estimation error.
The Fisher information matrix (FIM) associated with the unknown
parameter vector ${\boldsymbol{\zeta}}$ is given by
\begin{equation}
	{\mathbf{I}}({\bm{\zeta }}) = \mathbb{E}\left[ {{\left( {\frac{{\partial \ln p\left( {\tilde{\mathbf{y}};{\bm{\zeta }}} \right)}}{{\partial {\bm{\zeta }}}}} \right)}}{{\left( {\frac{{\partial \ln p\left( {\tilde{\mathbf{y}};{\bm{\zeta }}} \right)}}{{\partial {\bm{\zeta }}}}} \right)}^T} \right].
\end{equation}

After some algebraic manipulation, the FIM may be partitioned as
\begin{equation}
	{\bf I}({\boldsymbol{\zeta}})
	=
	\begin{bmatrix}
		{\bf I}_{rr} & {\bf I}_{ri} & {\bf I}_{rx} & {\bf I}_{ry}\\
		{\bf I}_{ir} & {\bf I}_{ii} & {\bf I}_{ix} & {\bf I}_{iy}\\
		{\bf I}_{xr} & {\bf I}_{xi} & {\bf I}_{xx} & {\bf I}_{xy}\\
		{\bf I}_{yr} & {\bf I}_{yi} & {\bf I}_{yx} & {\bf I}_{yy}
	\end{bmatrix},
\end{equation}
where each block element is defined as
\begin{equation}
	{\bf I}_{ab}
	=
	2\sigma_{n}^{-2}
	\Re\!\left(
	\sum_{l=1}^{L}
	{\bf D}_{l,a}^{H}
	\!\left(
	{\bf X}^{*}{\bf X}^{H}\!\otimes{\bf I}_{M}
	\right)
	{\bf D}_{l,b}
	\right),
\end{equation}
where ${\bf D}_{l,a}$ and ${\bf D}_{l,b}$ are the derivation matrices, with $\{a,b\}\in\{r,i,x,y\}$.

The derivative matrices are given by ${\bf D}_{l,r}
=
{\rm diag}\!\big({\bf d}_{l,r}\big)
\in\mathbb{C}^{K N_{\text{all}}\times K N_{\text{all}}},
{\bf D}_{l,i}
=
{\rm diag}\!\big({\bf d}_{l,i}\big)
\in\mathbb{C}^{K N_{\text{all}}\times K N_{\text{all}}},
{\bf D}_{l,x}
=
{\rm blkdiag}\!\big({\bf d}_{l,x,1},\ldots,{\bf d}_{l,x,K}\big)
\in\mathbb{C}^{M K\times K},
{\bf D}_{l,y}
=
{\rm blkdiag}\!\big({\bf d}_{l,y,1},\ldots,{\bf d}_{l,y,K}\big)
\in\mathbb{C}^{M K\times K}$
where the individual vectors are defined as
\begin{equation}
	{\bf d}_{l,r}
	=
	\!\left[
	e^{-j2\pi\frac{\tau_{1,11}}{LT}(l-1)},
	\ldots,
	e^{-j2\pi\frac{\tau_{K,MN_{m}}}{LT}(l-1)}
	\right]^T,
\end{equation}
\begin{equation}
	{\bf d}_{l,i}
	=
	\!\left[
	j\,e^{-j2\pi\frac{\tau_{1,11}}{LT}(l-1)},
	\ldots,
	j\,e^{-j2\pi\frac{\tau_{K,MN_{m}}}{LT}(l-1)}
	\right]^T,
\end{equation}
and 
\begin{equation}
	\big[{\bf d}_{l,x,k}\big]_{m}
	=
	\frac{-j2\pi}{LT}(l-1)
	H_{km}[l]
	\frac{x_{k}-x_{mn}^{\text{anc}}}
	{c\,\|\boldsymbol{\psi}_{k}-\boldsymbol{\psi}_{mn}^{\text{anc}}\|},
\end{equation}
\begin{equation}
	\big[{\bf d}_{l,y,k}\big]_{m}
	=
	\frac{-j2\pi}{LT}(l-1)
	H_{km}[l]
	\frac{y_{k}-y_{mn}^{\text{anc}}}
	{c\,\|\boldsymbol{\psi}_{k}-\boldsymbol{\psi}_{mn}^{\text{anc}}\|},
\end{equation}
for $m=1,\cdots,M$.

The derivation is given in Appendix A.

Finally, the CRLB matrix is obtained as
\begin{equation}
	{\bf C}({\boldsymbol{\zeta}})
	=
	\big[
	{\bf I}({\boldsymbol{\zeta}})
	\big]^{-1},
\end{equation}
with the CRLB being the diagonal elements of ${\bf C}({\boldsymbol{\zeta}})$.

\vspace{1ex}
\subsection{Complexity Analysis}

The computational complexity of the proposed framework is
categorized into two major components: the
EP phase and the
parameter extraction phase.

\textbf{1) EP Estimation:}
The dominant operations consist of the matrix multiplications and inversions in
\eqref{sigmao} and~\eqref{muo}. By precomputing ${\bf M}^{H}{\bf M}$ and exploiting
its block-sparse structure, the overall per-iteration complexity becomes
$\mathcal{O}(K^{3}L)+\mathcal{O}(MKPL)$. Meanwhile, the scalar variance updates have a
negligible cost, while the vector variance updates incur
$\mathcal{O}(MKL)$ complexity, which remains subordinate to that of the
matrix inversion.

\textbf{2) Parameter Extraction:}
For the OMP-based parameter extraction, which involves computing the
pseudo-inverse and correlation operations, the complexity of identifying
$K N_{\text{all}}$ active atoms is
$\mathcal{O}(K N_{\text{all}} N_{d} L)
+\mathcal{O}(K L \sum_{m=1}^{M} N_{m}^{2})$.
In the message-passing-based algorithm, updating
${\bf s}_{k,mn}$ requires
$\mathcal{O}(N_{\text{it}}^{\text{in}} K L
\sum_{m=1}^{M} N_{m}^{2})$ operations, while updating the scalar
parameters $\theta_{k,mn}$ and $z_{k,mn}$ contributes
$\mathcal{O}(N_{\text{it}}^{\text{in}} K L N_{\text{all}})$.

The localization refinement step incurs
$\mathcal{O}(K N_{\text{all}})$, whereas the delta method requires
$\mathcal{O}(KML(N_{\text{all}}+K^{2}))$.
Hence, the total complexity of the EP framework with
$N_{\text{it}}^{\text{out}}$ outer iterations can be expressed as
$\mathcal{O}(N_{\text{it}}^{\text{out}}K^{3}L)
+
\mathcal{O}(N_{\text{it}}^{\text{out}}MKPL)
+
\mathcal{O}(N_{\text{it}}^{\text{in}}N_{\text{it}}^{\text{out}}K L
\sum_{m=1}^{M} N_{m}^{2})
+
\mathcal{O}(N_{\text{it}}^{\text{in}}N_{\text{it}}^{\text{out}}K L N_{\text{all}})
+
\mathcal{O}(N_{\text{it}}^{\text{out}}KML(N_{\text{all}}+K^{2}))$. In contrast, employing the OMP-based extraction yields a total
complexity of $\mathcal{O}(N_{\text{it}}^{\text{out}}K^{3}L)
+
\mathcal{O}(N_{\text{it}}^{\text{out}}MKPL)
+
\mathcal{O}(N_{\text{it}}^{\text{out}}K N_{\text{all}} N_{d} L)
+
\mathcal{O}(N_{\text{it}}^{\text{out}}K L \sum_{m=1}^{M} N_{m}^{2})
+
\mathcal{O}(N_{\text{it}}^{\text{out}}KML(N_{\text{all}}+K^{2})).$

\section{Numerical Simulations}

\begin{figure*}[t]
	\centering
	\begin{subfigure}{0.32\textwidth}
		\centering
		\includegraphics[width=\linewidth]{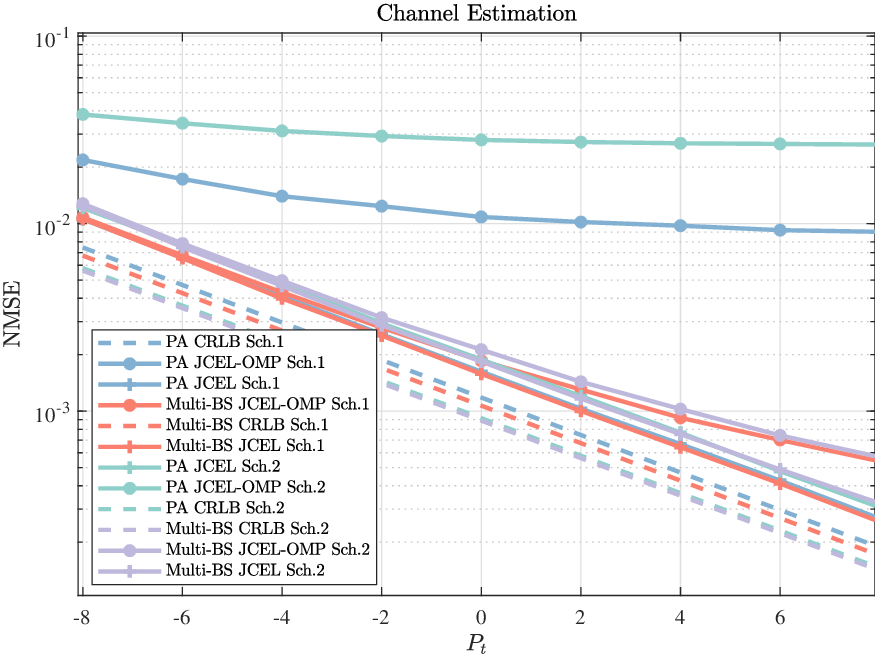}
		\caption{}
		\label{fig:1}
	\end{subfigure}
	\begin{subfigure}{0.32\textwidth}
		\centering
		\includegraphics[width=\linewidth]{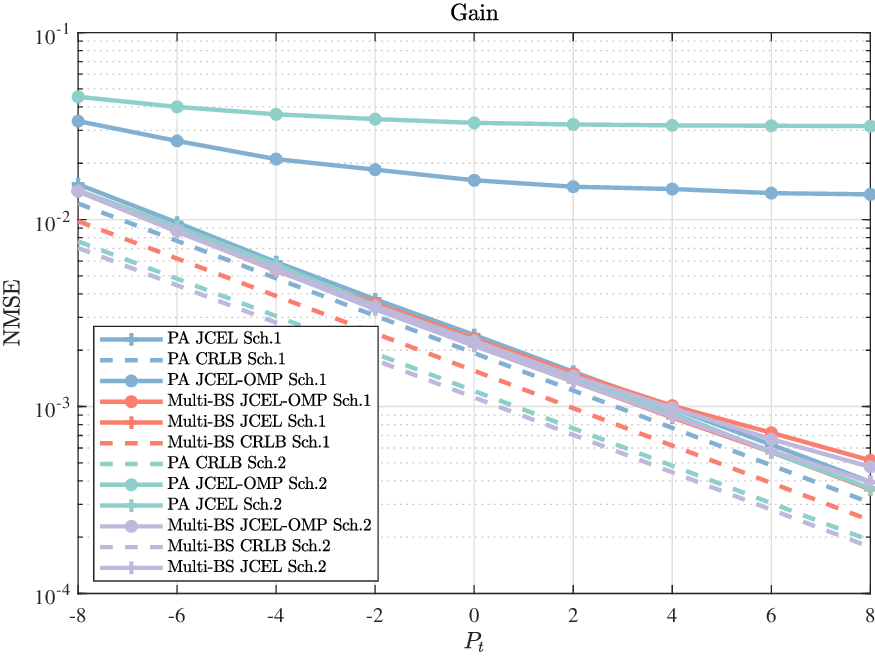}
		\caption{}
		\label{fig:2}
	\end{subfigure}
	\begin{subfigure}{0.32\textwidth}
		\centering
		\includegraphics[width=\linewidth]{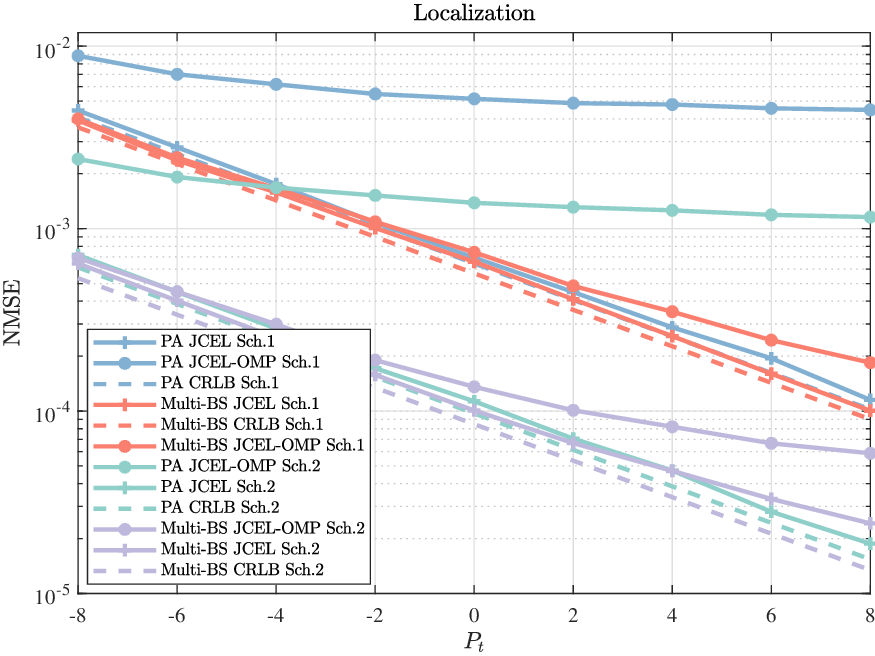}
		\caption{}
		\label{fig:3}
	\end{subfigure}
	
	\begin{subfigure}{0.32\textwidth}
		\centering
		\includegraphics[width=\linewidth]{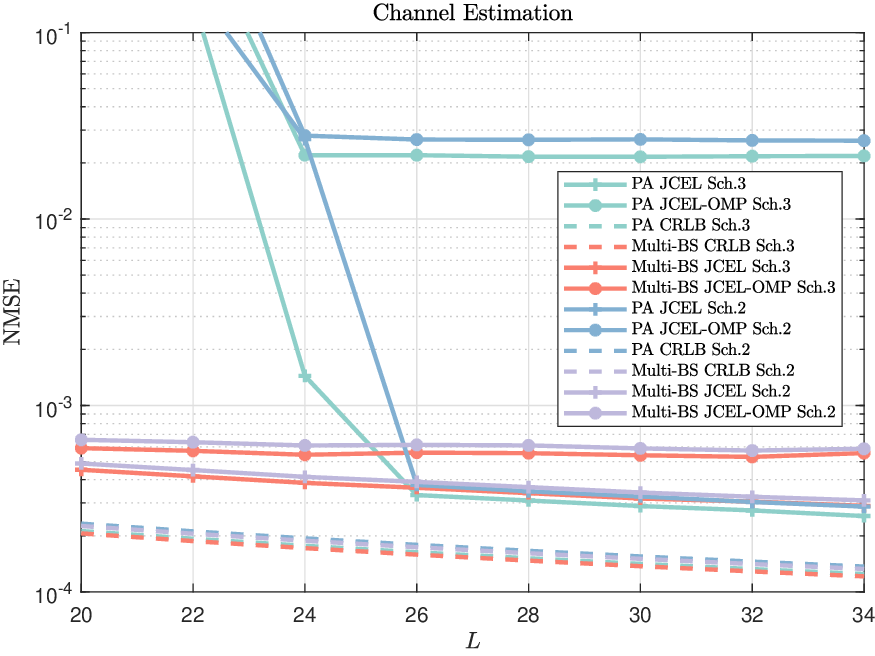}
		\caption{}
		\label{fig:4}
	\end{subfigure}
	\begin{subfigure}{0.32\textwidth}
		\centering
		\includegraphics[width=\linewidth]{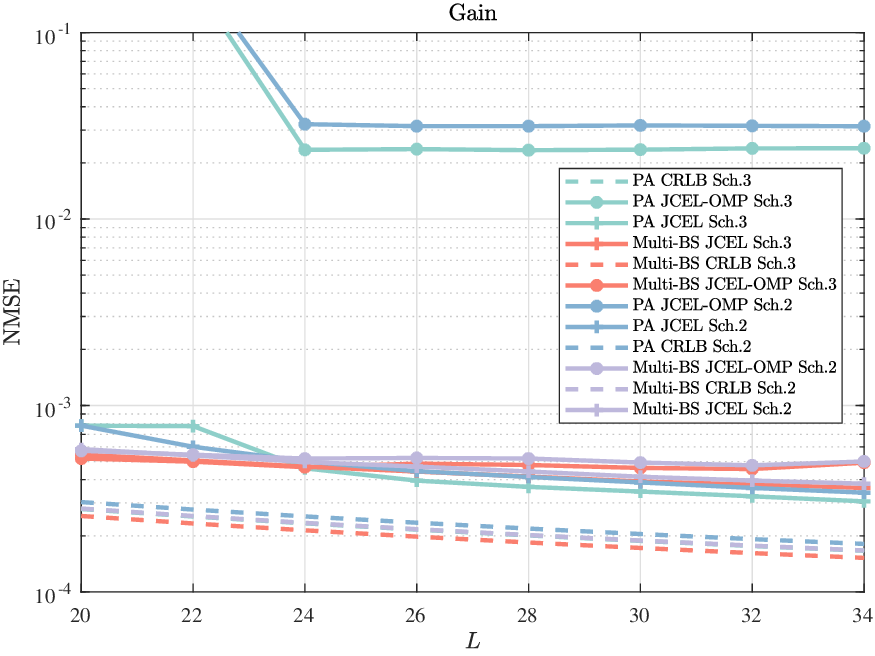}
		\caption{}
		\label{fig:5}
	\end{subfigure}
	\begin{subfigure}{0.32\textwidth}
		\centering
		\includegraphics[width=\linewidth]{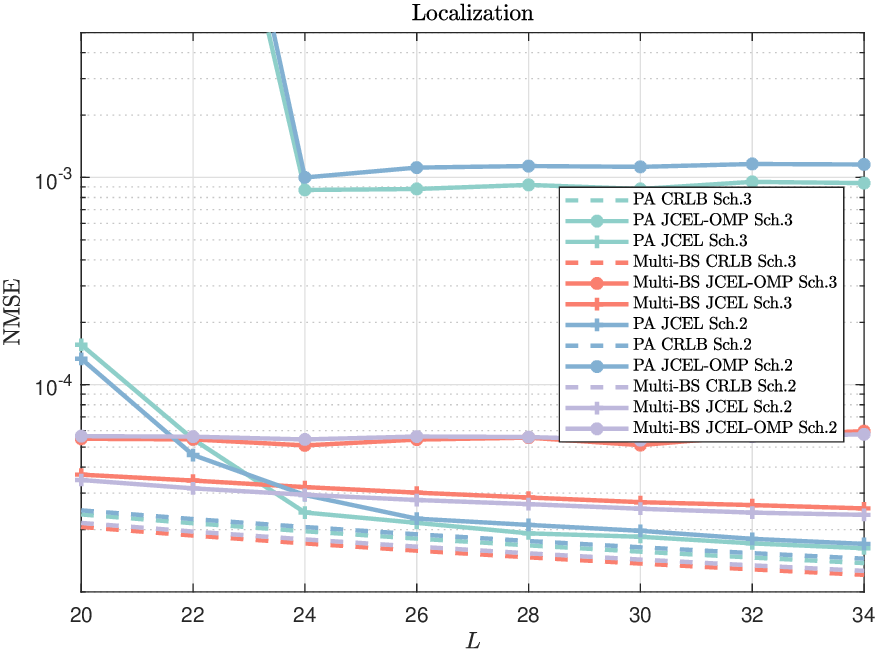}
		\caption{}
		\label{fig:6}
	\end{subfigure}
	
	\begin{subfigure}{0.32\textwidth}
		\centering
		\includegraphics[width=\linewidth]{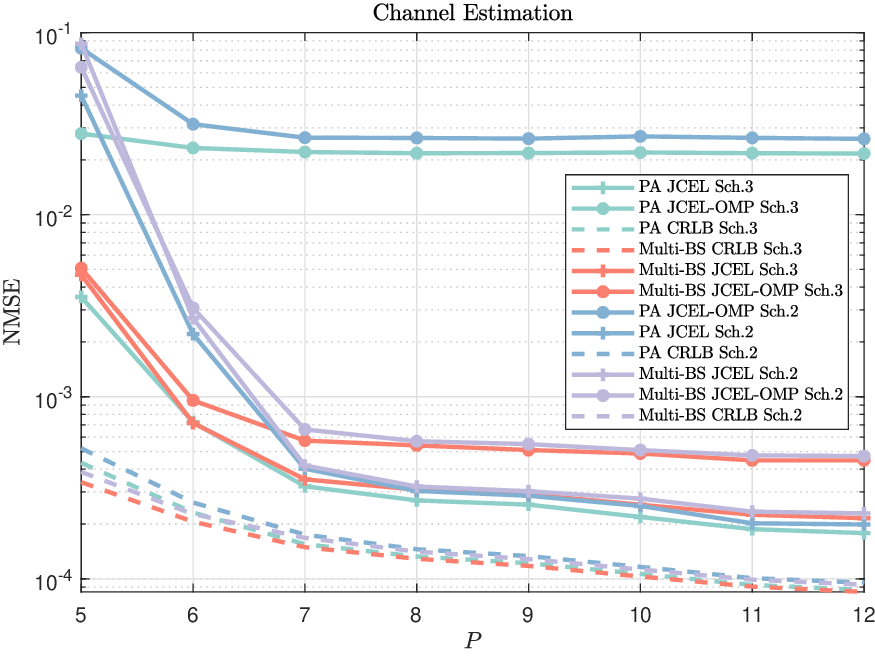}
		\caption{}
		\label{fig:7}
	\end{subfigure}
	\begin{subfigure}{0.32\textwidth}
		\centering
		\includegraphics[width=\linewidth]{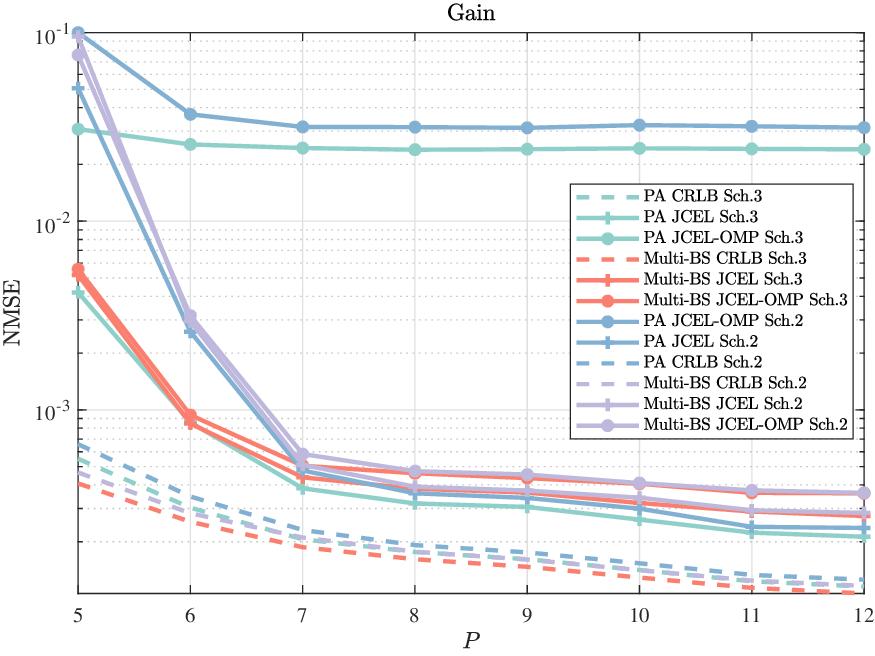}
		\caption{}
		\label{fig:8}
	\end{subfigure}
	\begin{subfigure}{0.32\textwidth}
		\centering
		\includegraphics[width=\linewidth]{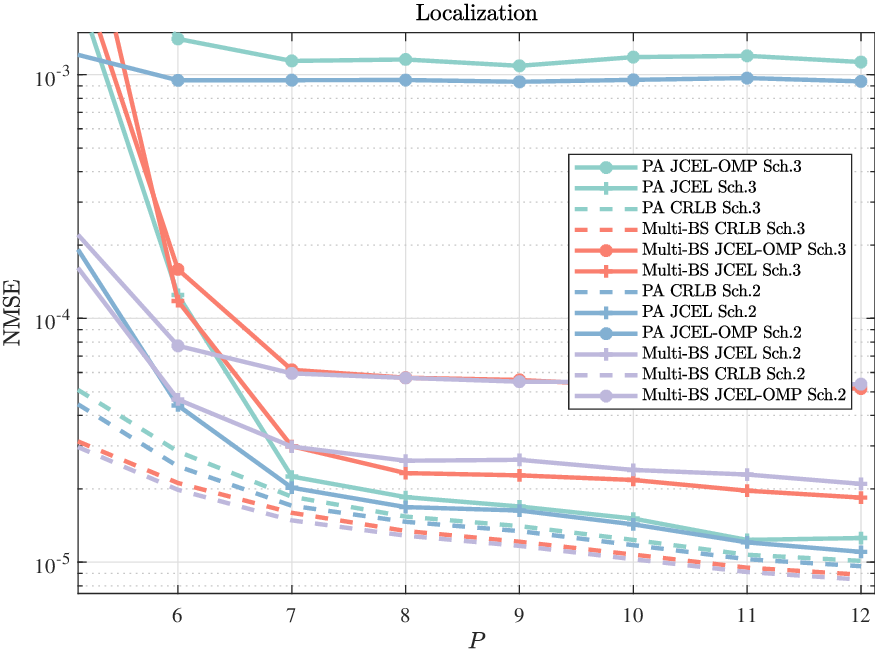}
		\caption{}
		\label{fig:9}
	\end{subfigure}
	
	\caption{
		Channel estimation, fading coefficient estimation, and localization performance under varying parameters:  
		(\subref{fig:1})–(\subref{fig:3}): transmit power $P_t$,  
		(\subref{fig:4})–(\subref{fig:6}): subcarrier count $L$, and  
		(\subref{fig:7})–(\subref{fig:9}): number of aggregated frames $P$.  
		``PA'' corresponds to the proposed PA-JCEL, whereas ``Multi-BS'' denotes multiple BS cooperative localization.
	}
	\label{fig:simall}
\end{figure*}

Numerical simulations are carried out to evaluate the performance of the proposed joint localization and channel–estimation framework for pinching-antenna (PA) systems, incorporating both OMP and message-passing-based delay extraction techniques. 
Unless otherwise specified, the system parameters follow Tab.~\ref{tab:simparams}.

\begin{table}[htb]
	\centering
	\caption{Simulation parameters.}
	\begin{tabular}{lcl}
		\hline\hline
		Parameter                     & Symbol                          & Value        \\
		\hline
		Central frequency             & $f_c$     & $28\,\mathrm{GHz}$      \\
		Bandwidth                 & $B$                           & $100\,\mathrm{MHz}$     \\
		Effective refractive factor         & $n_d$                              & 1.4      \\
		Total subcarriers             & $L$               & $32$                   \\
		Cyclic prefix point           & $L_{CP}$               & $16$                                     \\
		Gaussian noise power  & $N_0$               & $-90$ dBm \\
		Height of the waveguides  & $z^{anc}$               & $3$ m \\
		Maximum of inner iteration  & $N_\text{it}^\text{in}$               & $100$ \\
		Maximum number of outer EP iteration  & $N_\text{it}^\text{out}$               & $20$ \\
		\hline\hline
	\end{tabular}
	\label{tab:simparams}
\end{table}

\begin{table}[htb]
	\centering
	\caption{Positions of the PAs and users.}
	\begin{tabular}{lcl}
		\hline\hline
		Category                        & Coordinates \\ \hline
		PAs on waveguide 1              & $(-10,-5,3)$, $(-10,0,3)$, $(-10,5,3)$ \\
		PAs on waveguide 2              & $(-5,10,3)$, $(0,10,3)$, $(5,10,3)$ \\ 
		Users                           & $(2.85,1,0)$, $(3,-0.8,0)$, $(-2,2.3,0)$, $(1.5,-3,0)$ \\ 
		\hline\hline
	\end{tabular}
	\label{tab:posparams}
\end{table}

Since no existing PA-based localization systems are capable of jointly extracting the ToA information from multiple PAs residing on the same dielectric waveguide, we adopt an \textit{upper-bound} benchmark, i.e., the multi-BS cooperative localization. 
When only a single PA is activated on each waveguide, the PA system reduces exactly to this cooperative multi-BS case. 
Unlike the proposed PA based scheme, no delay superposition occurs and the estimation problem becomes substantially easier, resulting in more favorable theoretical bounds.  
The proposed iterative algorithm can also be applied to this benchmark scheme without modification.

\begin{figure}[htb]
	\centering
	\includegraphics[width=0.8\linewidth]{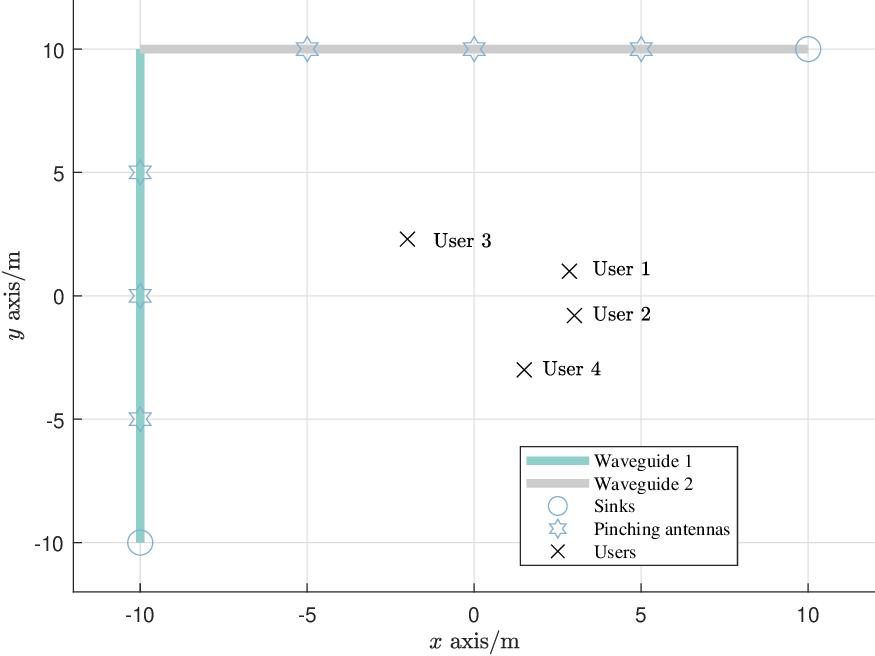}
	\caption{The spatial configuration of PAs and users.}
	\label{fig:Simconfig}
\end{figure}

The performance metrics are based on the normalized mean-square error (NMSE).  
The EP damping factor is fixed at 0.1, and the delay dictionary in OMP is set to contain 1000 atoms.  
The user/PA placement follows Table~\ref{tab:posparams} and Fig.~\ref{fig:Simconfig}. 
For the cooperative baseline, the number of BSs equals the number of PAs, and their locations coincide.  
The pilot matrix ${\mathbf X}$ is extracted from a matrix generated using the length $31$ Zadoff-Chu (ZC) sequence with the root indices $1$ to $31$.  
All users transmit with equal power, and the overall fading model is free-space without in-waveguide attenuation, i.e, $z_{kmn}^{i}=1$. In all simulations, the theoretical CRLBs are included for comparison.

Four simulation configurations are examined as follows.

\begin{itemize}
	\item \textbf{Scheme 1:} Only waveguide~1 is active, and only Users~1–3 are estimated ($M=1$, $N_1=3$, $K=3$).
	\item \textbf{Scheme 2:} Both waveguides are active, and all four users are estimated ($M=2$, $N_1=N_2=3$, $K=4$).
	\item \textbf{Scheme 3:} Both waveguides are active, but only Users~1–3 are estimated ($M=2$, $N_1=N_2=3$, $K=3$).
	\item \textbf{Scheme 4:} Only waveguide~1 is active, but all four users are estimated ($M=1$, $N_1=3$, $K=4$).
\end{itemize}

In Schemes~1 and~2, the subcarrier count and frame aggregation level are fixed at $L=32$ and $P=8$, while the transmit power $P_t$ is varied.  
In Scheme~3, the parameters $L$ and $P$ are varied, while $P_t$ is fixed at $8$ dBm.

The estimation NMSE of the channel coefficient ${\bf H}[l]$, the complex fading coefficient $z_{kmn}$ and the location $\bm \psi_{k}$, $k=1,\cdots, K$, $m=1,\cdots,M$, $n=1,\cdots,N$, $l=1,\cdots,L$ are evaluated in different schemes, and the results are given in Fig. \ref{fig:simall}.

Fig.~\ref{fig:simall} (\subref{fig:1})-\ref{fig:simall} (\subref{fig:3}) illustrate the RMSE versus varied transmit power. When the transmit power increases, the performance of JCEL steadily improves in both the multiple BS localization case and the PA localization case. Moreover, it may be observed that both the theoretical and simulated performance gaps between the PA-based scheme and the multiple BS cooperative baseline remain marginal across all metrics. 
All the estimation accuracy obtained through message-passing delay extraction closely follows their cooperative counterparts.  
Specifically, the localization accuracy exhibits less than 1 dB deviation from the CRLB, showing moderate localization performance. 
Given that the PA architecture dramatically reduces the required number of RF chains (from $N_{\mathrm{all}}$ to $M$), the PA-based JCEL framework is appealing from a cost-efficiency standpoint for uplink localization.

Fig.~\ref{fig:simall} (\subref{fig:1})-\ref{fig:simall} (\subref{fig:3}) also reveal that, in the cooperative multi-BS setting, the OMP-based delay extraction performs comparably to message passing and remains close to the CRLB.  
However, under the PA based JCEL, OMP suffers from an early error floor.  
This degradation may stem from the superimposed multi-path delays observed at the waveguide output in the PA case, which reduce the effective sparsity and amplify mutual interference between dictionary atoms, compared to the multiple BS localization case where only a single delay without superposition could be extracted and the sparsity level is therefore high enough.  
In contrast, the message-passing off-grid delay estimator retains robust and maintains competitive performance in both cases.

The impact of subcarrier count $L$ is illustrated in Fig.~\ref{fig:simall} (\subref{fig:4})-\ref{fig:simall} (\subref{fig:6}).  With the increasing of samples $L$, the estimation performance improves accordingly.
With insufficient $L$ under given $P_t$, the delay superposition in the PA becomes difficult to fully resolve, leading to degraded estimation accuracy relative to the cooperative baseline, especially in localization and channel estimation.  
Once $L$ becomes sufficiently large, the estimation performance using PA saturates and approaches that of multi-BS localization. Moreover, the message passing based JCEL always outperforms the OMP based one when $L$ is sufficiently large. Similarly, as shown in Fig.~\ref{fig:simall} (\subref{fig:7})-\ref{fig:simall} (\subref{fig:9}), increasing the number of aggregated frames $P$ could also provide moderate improvements for both architectures, by enhancing the effective SNR of the output in the linear processing stage of EP.

\begin{figure}[htb]
	\centering
	\includegraphics[width=0.8\linewidth]{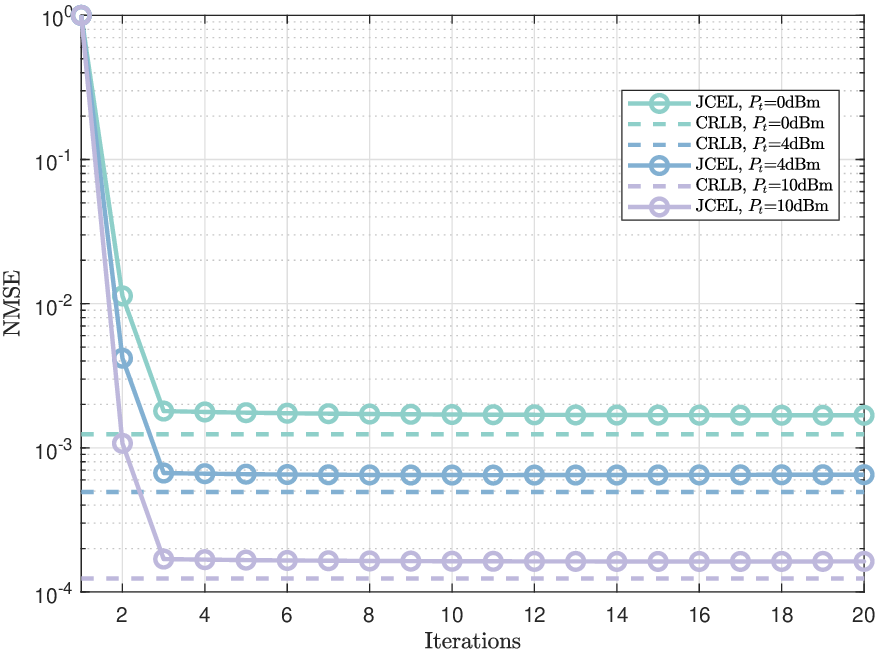}
	\caption{
		Convergence behavior of the EP outer iteration under Scheme 4 for different transmit powers $P_t$, where the $y$ axis denotes the NMSE of the channel estimation.
	}
	\label{fig:Itersim}
\end{figure}

The influence of in-waveguide loss is also examined in our simulations.  
Following the model in~\cite{PAloss}, the attenuation induced by the dielectric waveguide is characterized as an exponential decay of the complex gain, expressed as
\begin{equation}
	z_{mn}^{i}
	= 
	\sqrt{
		\exp\!\left(
		-k_{\mathrm{loss}}
		\left\|
		\boldsymbol{\psi}_{m}^{\mathrm{anc}}
		-
		\boldsymbol{\psi}_{mn}^{\mathrm{anc}}
		\right\|
		\right)
	}.
\end{equation}

To quantify the effect of waveguide attenuation, we evaluate the NMSE of channel estimation and localization under Scheme~1 for three representative attenuation constants:  
(i) $k_{\mathrm{loss}} = 0.01$, corresponding to ultra–low-loss dielectric materials,  
(ii) $k_{\mathrm{loss}} = 0.1$, associated with moderate-loss pure-ceramic waveguides~\cite{PAloss2} and
(ii) $k_{\mathrm{loss}} = 0.05$, the intermediate scheme. 

In all tested scenarios, the localization performance remains closely aligned with the theoretical CRLB, confirming the robustness of the proposed JCEL framework.  
The dominant effect of in-waveguide loss is the reduction in the effective SNR.  
As the attenuation increases, the received signal power decreases accordingly, which results in a degradation of both channel estimation and localization accuracy.  
For ultra–low-loss materials, the resulting performance degradation is negligible (typically below 1 dB), whereas for pure-ceramic materials with higher loss characteristics, the degradation becomes non-negligible and must be accounted for in practical system design.

\begin{figure}[t]
	\centering
	\begin{subfigure}{0.45\textwidth}
		\centering
		\includegraphics[width=\linewidth]{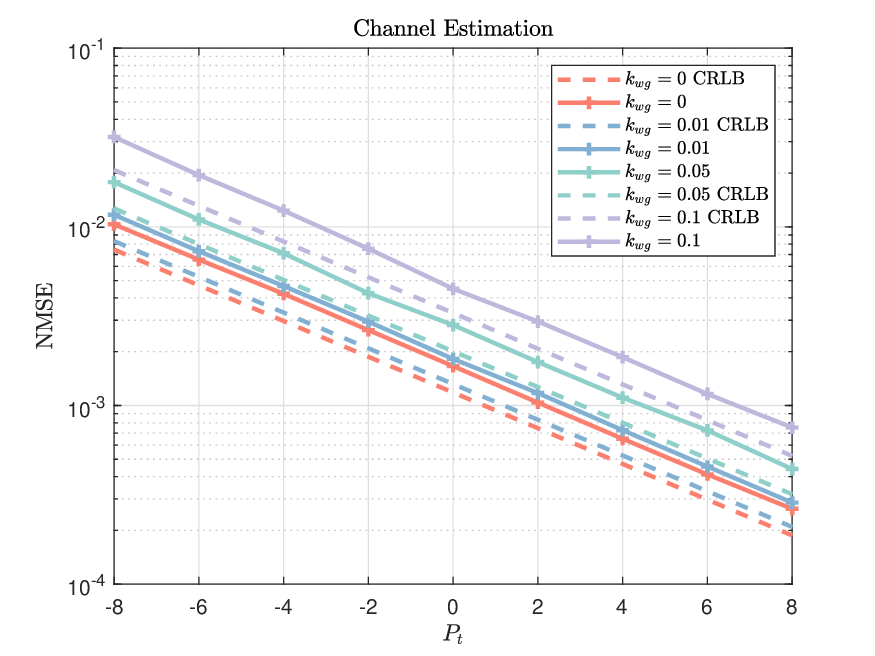}
		\caption{}
		\label{fig:kwgce}
	\end{subfigure}
	\begin{subfigure}{0.45\textwidth}
		\centering
		\includegraphics[width=\linewidth]{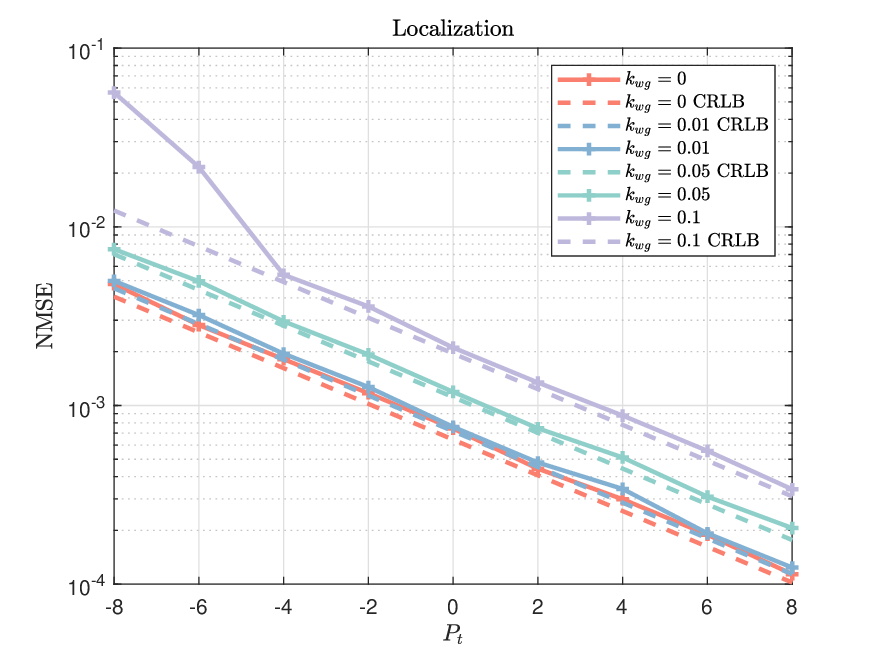}
		\caption{}
		\label{fig:kwggain}
	\end{subfigure}
	
	\caption{
		(\subref{fig:kwgce}): Channel estimation and (\subref{fig:kwggain}): localization performance of the JCEL under different in-waveguide attenuation.
	}
	\label{fig:simkwg}
\end{figure}

Finally, Fig.~\ref{fig:Itersim} illustrates the empirical convergence speed of the outer EP loop, where the NMSE of the channel coefficient ${\bf H}[l]$ under Scheme 4 versus iteration is used for benchmark.
Across several transmit powers tested, convergence to a stable fixed point close to the theoretical bound occurs within approximately five iterations, indicating the numerical stability of the proposed inference mechanism.

\section{Conclusion}

In this paper, we investigated the problem of joint channel estimation and localization for uplink OFDM based PASS. 
A comprehensive signal model was developed, capturing both the transmission structure and the inherent coupling between the channel state information and the user position. 
Building upon this model, an EP inference framework was formulated, within which two delay extraction modules, a dictionary-based OMP and an off-grid BP-VI estimator, were incorporated to retrieve the delay components.
The corresponding CRLBs were analytically derived to characterize the fundamental estimation limits of the PASS architecture, and extensive numerical simulations validated the effectiveness of the proposed approach, demonstrating that the low-cost PA-JCEL scheme can achieve performance remarkably close to that of a full–RF-chain multi-BS cooperative localization system, despite requiring substantially fewer hardware resources, thereby establishing PA-enabled JCEL as a promising and cost-efficient technology for future large-scale localization scenarios.

\appendix[Derivation of the FIM]

Let 
\(
{\mathbf{\tilde y}}
=
\left[
{\mathbf{\tilde y}}_1^{T},
\ldots,
{\mathbf{\tilde y}}_L^{T}
\right]^{T}
\)
and
\(
{\mathbf{\tilde h}}
=
\left[
{\mathbf{\tilde h}}_1^{T},
\ldots,
{\mathbf{\tilde h}}_L^{T}
\right]^{T}.
\)
Based on the signal model, the corresponding log likelihood function can be written as
\begin{equation}
\begin{split}
	\label{eq:loglike}
	&\ln p\!\left({\mathbf{\tilde y}};{\mathbf{\zeta}}\right)
	= 
	\text{const}
	-\\
	&\sum_{l=1}^{L}
	\frac{1}{\sigma_{n}^{2}}
	\left(
	{\mathbf{\tilde y}}_{l}
	-
	\left({\mathbf{X}}^{T}\!\otimes\!{\mathbf{I}}_{M}\right)
	{\mathbf{\tilde h}}_{l}
	\right)^{H}
	\left(
	{\mathbf{\tilde y}}_{l}
	-
	\left({\mathbf{X}}^{T}\!\otimes\!{\mathbf{I}}_{M}\right)
	{\mathbf{\tilde h}}_{l}
	\right).
\end{split}
\end{equation}

Taking the Wirtinger derivatives of~\eqref{eq:loglike} yields  
\begin{equation}\label{eq:dlogdh}
\begin{split}
	&\frac{\partial \ln p}{\partial {\mathbf{\tilde h}}_{l}}
	=
	\frac{1}{\sigma_{n}^{2}}
	\left({\mathbf{X}}^{H}\!\otimes\!{\mathbf{I}}_{M}\right)
	\left(
	{\mathbf{\tilde y}}_{l}^{*}
	-
	\left({\mathbf{X}}^{H}\!\otimes\!{\mathbf{I}}_{M}\right)
	{\mathbf{\tilde h}}_{l}^{*}
	\right)\\
	&=
	\frac{1}{\sigma_{n}^{2}}
	\left({\mathbf{X}}^{H}\!\otimes\!{\mathbf{I}}_{M}\right)
	{\mathbf{n}}_{l}^{*},
\end{split}
\end{equation}
\begin{equation}
	\label{eq:dlogdhc}
	\begin{split}
	&\frac{\partial \ln p}{\partial {\mathbf{\tilde h}}_{l}^{*}}
	=
	\frac{1}{\sigma_{n}^{2}}
	\left({\mathbf{X}}^{T}\!\otimes\!{\mathbf{I}}_{M}\right)
	\left(
	{\mathbf{\tilde y}}_{l}
	-
	\left({\mathbf{X}}^{T}\!\otimes\!{\mathbf{I}}_{M}\right)
	{\mathbf{\tilde h}}_{l}
	\right)\\
	&=
	\frac{1}{\sigma_{n}^{2}}
	\left({\mathbf{X}}^{T}\!\otimes\!{\mathbf{I}}_{M}\right)
	{\mathbf{n}}_{l},
\end{split}
\end{equation}
where \({\mathbf{n}}_{l}\) denotes circularly symmetric AWGN with variance \(\sigma_{n}^{2}\).  
Noise samples corresponding to different subcarriers are uncorrelated, i.e.,
\begin{equation}
	\mathbb{E}\!\left({\mathbf{n}}_{l_{1}}{\mathbf{n}}_{l_{2}}^{H}\right)
	=
	\mathbf{0}, \quad l_{1}\neq l_{2},
\end{equation}
\begin{equation}
	\mathbb{E}\!\left({\mathbf{n}}_{l}{\mathbf{n}}_{l}^{H}\right)
	=
	\sigma_{n}^{2}\mathbf{I}_{MP}.
\end{equation}
  
Using the chain rule, we first have
\begin{equation}
	\frac{\partial \ln p}{\partial {\mathbf{\zeta}}}
	=
	\sum_{l=1}^{L}
	\left[
	\left(\frac{\partial {\mathbf{\tilde h}}_{l}}{\partial {\mathbf{\zeta}}}\right)^{T}
	\frac{\partial \ln p}{\partial {\mathbf{\tilde h}}_{l}}
	+
	\left(\frac{\partial {\mathbf{\tilde h}}_{l}^{*}}{\partial {\mathbf{\zeta}}}\right)^{T}
	\frac{\partial \ln p}{\partial {\mathbf{\tilde h}}_{l}^{*}}
	\right].
\end{equation}

Since Fisher information matrix (FIM) is defined as
\begin{equation}
	{\mathbf{I}}({\mathbf{\zeta}})
	=
	\mathbb{E}
	\!\left[
	\left(\frac{\partial \ln p}{\partial {\mathbf{\zeta}}}\right)
	\left(\frac{\partial \ln p}{\partial {\mathbf{\zeta}}}\right)^{T}
	\right],
\end{equation}
substituting~\eqref{eq:dlogdh}–\eqref{eq:dlogdhc} and using the noise independence property yields
\begin{equation}
	\label{eq:FIMfinal}
	{\mathbf{I}}({\mathbf{\zeta}})
	=
	2\sigma_{n}^{-2}
	\sum_{l=1}^{L}
	\Re
	\!\left[
	\left(\frac{\partial {\mathbf{\tilde h}}_{l}}{\partial {\mathbf{\zeta}}}\right)^{T}
	\left({\mathbf{X}}^{*}{\mathbf{X}}^{H} \otimes {\mathbf{I}}_{M}\right)
	\left(\frac{\partial {\mathbf{\tilde h}}_{l}}{\partial {\mathbf{\zeta}}}\right)
	\right].
\end{equation}

The remaining task is to evaluate 
\(
\frac{\partial {\mathbf{\tilde h}}_{l}}{\partial {\mathbf{\zeta}}}
\)
for all parameters of interest and assemble the result into matrix form, which completes the derivation of the FIM.

\ifCLASSOPTIONcaptionsoff
  \newpage
\fi

\bibliography{IEEEabrv,ref} 

@article{VAMP,
  title={Vector approximate message passing},
  author={Rangan, Sundeep and Schniter, Philip and Fletcher, Alyson K},
  journal={IEEE Trans. Inf. Theory},
  volume={65},
  number={10},
  pages={6664--6684},
  year={2019},
  month={May},
  publisher={IEEE}
}

@article{PA2,
  title={Pinching-Antenna Systems ({PASS}): Architecture Designs, Opportunities, and Outlook},
  author={Liu, Yuanwei and Wang, Zhaolin and Mu, Xidong and Ouyang, Chongjun and Xu, Xiaoxia and Ding, Zhiguo},
  year={2025},
  note={\textit{arXiv:2501.18409}}
}

@ARTICLE{PA1,
  author={Ding, Zhiguo and Schober, Robert and Vincent Poor, H.},
  journal={IEEE Trans. Commun.}, 
  title={Flexible-Antenna Systems: A Pinching-Antenna Perspective}, 
  year={2025},
  volume={73},
  number={10},
  pages={9236-9253},
  mon={Oct.}}

@article{PA3,
  title={Pinching antennas: Principles, applications and challenges},
  author={Yang, Zheng and Wang, Ning and Sun, Yanshi and Ding, Zhiguo and Schober, Robert and Karagiannidis, George K and Wong, Vincent W S and Dobre, Octavia A},
  year={2025},
  note={\textit{arXiv:2501.10753}}
}

@article{PARate,
  author={Xu, Yanqing and Ding, Zhiguo and Karagiannidis, George K},
  journal={IEEE Wireless Commun. Lett.},
  title={Rate Maximization for Downlink Pinching-Antenna Systems},
  year={2025},
  volume={14},
  number={5},
  pages={1431-1435},
  month={May.}
}

@article{PARate2,
  author={Zhou, Ziwu and Yang, Zheng and Chen, Gaojie and Ding, Zhiguo},
  journal={IEEE Wireless Commun. Lett.},
  title={Sum-Rate Maximization for {NOMA}-Assisted Pinching-Antenna Systems},
  year={2025},
  volume={14},
  number={9},
  pages={2728-2732},
  month={Sept.}
}

@article{PABF1,
  title={Downlink beamforming with pinching-antenna assisted {MIMO} systems},
  author={Bereyhi, Ali and Asaad, Saba and Ouyang, Chongjun and Ding, Zhiguo and Poor, H Vincent},
  year={2025},
  note={\textit{arXiv:2502.01590}}
}

@article{PABF2,
  title={Modeling and beamforming optimization for pinching-antenna systems},
  author={Wang, Zhaolin and Ouyang, Chongjun and Mu, Xidong and Liu, Yuanwei and Ding, Zhiguo},
  year={2025},
  note={\textit{arXiv:2502.05917}}
}

@article{PABFZY,
  title={Two-Dimensional Pinching-Antenna Systems: Modeling and Beamforming Design},
  author={Zhong, Yuan and Xiao, Yue and Li, Yijia and Chen, Hao and Lei, Xianfu and Fan, Pingzhi},
  year={2025},
  note={\textit{arXiv:2511.09207}}
}

@article{PAPLSZY,
  author={Zhong, Yuan and Chen, Jiangong and Xiao, Yue and Yang, Shuaixin and Lei, Xianfu and Gao, Yulan and Xiao, Ming},
  journal={IEEE Wireless Commun. Lett.},
  title={Physical Layer Security for Pinching-Antenna Systems via Index and Directional Modulation},
  year={2025},
  pages={},
  note={early access}
}

@article{PASensing1,
  title={Wireless Sensing via Pinching-Antenna Systems},
  author={Wang, Zhaolin and Ouyang, Chongjun and Liu, Yuanwei and Nallanathan, Arumugam},
  year={2025},
  note={\textit{arXiv:2505.15430}}
}

@article{PASensing2,
  title={Pinching Antenna System for Integrated Sensing and Communications},
  author={Li, Haochen and Zhong, Ruikang and Lei, Jiayi and Liu, Yuanwei},
  year={2025},
  note={\textit{arXiv:2508.19540}}
}

@article{PASensing3,
  title={Integrated sensing and communications for pinching-antenna systems ({PASS})},
  author={Zhang, Zheng and Wang, Zhaolin and Mu, Xidong and He, Bingtao and Chen, Jian and Liu, Yuanwei},
  year={2025},
  note={\textit{arXiv:2504.07709}}
}

@article{PASensing4,
  title={Cram{\'e}r-Rao Bounds for Integrated Sensing and Communications in Pinching-Antenna Systems},
  author={Bozanis, Dimitrios and Papanikolaou, Vasilis K and Tegos, Sotiris A and Karagiannidis, George K},
  year={2025},
  note={\textit{arXiv:2505.01333}}
}

@article{PASensing5,
  title={Rate region of {ISAC} for pinching-antenna systems},
  author={Ouyang, Chongjun and Wang, Zhaolin and Liu, Yuanwei and Ding, Zhiguo},
  year={2025},
  note={\textit{arXiv:2505.10179}}
}

@article{PASensing6,
  title={Pinching-Antenna System-Assisted Localization: A Stochastic Geometry Perspective},
  author={He, Jiajun and Mu, Xidong and Ngo, Hien Quoc and Matthaiou, Michail},
  year={2025},
  note={\textit{arXiv:2511.15444}}
}

@article{PASensing7,
  title={Pinching-Antenna Systems ({PASS})-based Indoor Positioning},
  author={Zhang, Yaoyu and Sun, Xin and Wang, Jun and Hou, Tianwei and Li, Anna and Liu, Yuanwei and Nallanathan, Arumugam},
  year={2025},
  note={\textit{arXiv:2508.08185}}
}

@article{PASensing8,
  title={Pinching-Antenna Assisted Sensing: A Bayesian Cram{\'e}r-Rao Bound Perspective},
  author={Jiang, Hao and Ouyang, Chongjun and Wang, Zhaolin and Liu, Yuanwei and Nallanathan, Arumugam and Ding, Zhiguo},
  year={2025},
  note={\textit{arXiv:2510.09137}}
}

@article{PAOFDM,
  title={{OFDMA} for Pinching Antenna Systems},
  author={Oikonomou, Thrassos K and Tegos, Sotiris A and Diamantoulakis, Panagiotis D and Liu, Yuanwei and Karagiannidis, George K},
  year={2025},
  note={\textit{arXiv:2505.19902}}
}

@article{PAOFDM2,
  author={Xiao, Jian and Wang, Ji and Zeng, Ming and Liu, Yuanwei and Karagiannidis, George K},
  journal={IEEE Wireless Commun. Lett.},
  title={Frequency-Selective Modeling and Analysis for {OFDM}-Integrated Wideband Pinching-Antenna Systems},
  year={2025},
  volume={14},
  number={11},
  pages={3500--3504},
  month={Nov.}
}

@article{FABF1,
  author={Qin, Haoran and Chen, Wen and Li, Zhendong and Wu, Qingqing and Cheng, Nan and Chen, Fangjiong},
  journal={IEEE Wireless Commun. Lett.},
  title={Antenna Positioning and Beamforming Design for Fluid Antenna-Assisted Multi-User Downlink Communications},
  year={2024},
  volume={13},
  number={4},
  pages={1073--1077},
  month={Apr.}
}

@article{FABF2,
  author={Ma, Wenyan and Zhu, Lipeng and Zhang, Rui},
  journal={IEEE Commun. Lett.},
  title={Multi-Beam Forming With Movable-Antenna Array},
  year={2024},
  volume={28},
  number={3},
  pages={697--701},
  month={Mar.}
}

@article{FAISAC,
  author={Lou, Xingliang and Xia, Wenchao and Zhu, Yongxu and Wong, Kai-Kit and Chae, Chan-Byoung},
  journal={IEEE Trans. Cognit. Commun. Networking},
  title={Multi-Target Beamforming Optimization for Fluid Antenna-Enabled Multi-Static {ISAC}},
  year={2025},
  pages={},
  note={early access}
}

@article{FAISAC2,
  author={Lyu, Wanting and Yang, Songjie and Xiu, Yue and Zhang, Zhongpei and Assi, Chadi and Yuen, Chau},
  journal={IEEE Trans. Wireless Commun.},
  title={Movable Antenna Enabled Integrated Sensing and Communication},
  year={2025},
  volume={24},
  number={4},
  pages={2862--2875},
  month={Apr.}
}

@article{FAPLS,
  author={Rostami Ghadi, Farshad and Wong, Kai-Kit and López-Martínez, F Javier and New, Wee Kiat and Xu, Hao and Chae, Chan-Byoung},
  journal={IEEE Trans. Wireless Commun.},
  title={Physical Layer Security Over Fluid Antenna Systems: Secrecy Performance Analysis},
  year={2024},
  volume={23},
  number={12},
  pages={18201--18213},
  month={Dec.}
}

@article{FAIM,
  author={Zhu, Jing and Chen, Gaojie and Gao, Pengyu and Xiao, Pei and Lin, Zihuai and Quddus, Atta Ul},
  journal={IEEE Trans. Wireless Commun.},
  title={Index Modulation for Fluid Antenna-Assisted {MIMO} Communications: System Design and Performance Analysis},
  year={2024},
  volume={23},
  number={8},
  pages={9701--9713},
  month={Aug.}
}

@article{FACC,
  author={Liu, Min and Xiao, Yue and Zhang, Lechen and Yang, Shuaixin and Wu, Chaowu and Lei, Xia},
  journal={IEEE Trans. Veh. Technol.},
  title={Index Modulation for Covert Transmission in Continuous-Trajectory Fluid Antenna Systems},
  year={2025},
  pages={1--6},
  note={early access}
}

@article{FACE1,
  author={New, Wee Kiat and Wong, Kai-Kit and Xu, Hao and Rostami Ghadi, Farshad and Murch, Ross and Chae, Chan-Byoung},
  journal={IEEE Trans. Wireless Commun.},
  title={Channel Estimation and Reconstruction in Fluid Antenna System: Oversampling is Essential},
  year={2025},
  volume={24},
  number={1},
  pages={309--322},
  month={Jan.}
}

@article{MACE1,
  author={Ma, Wenyan and Zhu, Lipeng and Zhang, Rui},
  journal={IEEE Commun. Lett.},
  title={Compressed Sensing Based Channel Estimation for Movable Antenna Communications},
  year={2023},
  volume={27},
  number={10},
  pages={2747--2751},
  month={Oct.}
}

@article{VALSE,
  title={Variational Bayesian inference of line spectra},
  author={Badiu, Mihai-Alin and Hansen, Thomas Lundgaard and Fleury, Bernard Henri},
  journal={IEEE Trans. Signal Process.},
  volume={65},
  number={9},
  pages={2247--2261},
  year={2017},
  month={May.},
  publisher={IEEE}
}

@article{VMP,
  title={Variational message passing},
  author={Winn, John and Bishop, Christopher M and Jaakkola, Tommi},
  journal={J. Mach. Learn. Res.},
  volume={6},
  number={4},
  month={Apr.},
  year={2005}
}

@book{PAML,
  title={Pattern recognition and machine learning},
  author={Bishop, Christopher M and Nasrabadi, Nasser M},
  year={2006},
  publisher={Springer}
}

@article{PAloss,
  author={Tyrovolas, Dimitrios and Tegos, Sotiris A and Diamantoulakis, Panagiotis D and Ioannidis, Sotiris and Liaskos, Christos K and Karagiannidis, George K},
  journal={IEEE Trans. Cognit. Commun. Networking},
  title={Performance Analysis of Pinching-Antenna Systems},
  year={2025},
  note={early access}

}

@article{PAloss2,
  author={Wang, Kaidi and Ding, Zhiguo and Schober, Robert},
  journal={IEEE Wireless Commun. Lett.},
  title={Antenna Activation for {NOMA}-Assisted Pinching-Antenna Systems},
  year={2025},
  volume={14},
  number={5},
  month={May.},
  pages={1526--1530}
}

@article{FA1,
  author={Wong, Kai-Kit and Shojaeifard, Arman and Tong, Kin-Fai and Zhang, Yangyang},
  journal={IEEE Trans. Wireless Commun.},
  title={Fluid Antenna Systems},
  year={2021},
  volume={20},
  number={3},
  month={Mar.},
  pages={1950--1962}
}

@article{MA1,
  author={Zhu, Lipeng and Ma, Wenyan and Zhang, Rui},
  journal={IEEE Commun. Mag.},
  title={Movable Antennas for Wireless Communication: Opportunities and Challenges},
  year={2024},
  volume={62},
  number={6},
  month={Jun.},
  pages={114--120}
}

@article{PAPLS1,
  author={Wang, Kaidi and Ding, Zhiguo and Al-Dhahir, Naofal},
  journal={IEEE Wireless Commun. Lett.},
  title={Pinching-Antenna Systems for Physical Layer Security},
  year={2025},
  note={early access}
}

@article{Loc1,
  author={Patwari, N and Ash, J N and Kyperountas, S and Hero, A O and Moses, R L and Correal, N S},
  journal={IEEE Signal Process. Mag.},
  title={Locating the nodes: cooperative localization in wireless sensor networks},
  year={2005},
  volume={22},
  number={4},
  month={Jul.},
  pages={54--69}
}

@article{EP1,
  title={Expectation propagation for approximate {B}ayesian inference},
  author={Minka, Thomas P},
  note={\textit{arXiv:1301.2294}},
}

@article{EP2,
  title={Expectation propagation detection for high-order high-dimensional {MIMO} systems},
  author={Cespedes, Javier and Olmos, Pablo M and S{\'a}nchez-Fern{\'a}ndez, Matilde and Perez-Cruz, Fernando},
  journal={IEEE Trans. Commun.},
  volume={62},
  number={8},
  pages={2840--2849},
  year={2014},
  month={Aug.},
  publisher={IEEE}
}

@ARTICLE{6G1,
  author={Wang, Cheng-Xiang and You, Xiaohu and Gao, Xiqi and Zhu, Xiuming and Li, Zixin and Zhang, Chuan and Wang, Haiming and Huang, Yongming and Chen, Yunfei and Haas, Harald and Thompson, John S. and Larsson, Erik G. and Renzo, Marco Di and Tong, Wen and Zhu, Peiying and Shen, Xuemin and Poor, H. Vincent and Hanzo, Lajos},
  journal={IEEE Commun. Surv. Tutorials}, 
  title={On the Road to 6G: Visions, Requirements, Key Technologies, and Testbeds}, 
  year={2023},
  volume={25},
  number={2},
  pages={905-974},
  month={Feb.}
}

@ARTICLE{6G2,
  author={Xiao, Yue and Ye, Ziqiang and Wu, Mingming and Li, Haoyun and Xiao, Ming and Alouini, Mohamed-Slim and Al-Hourani, Akram and Cioni, Stefano},
  journal={IEEE J. Sel. Areas Commun.}, 
  title={Space-Air-Ground Integrated Wireless Networks for 6G: Basics, Key Technologies, and Future Trends}, 
  year={2024},
  volume={42},
  number={12},
  pages={3327-3354},
  month={Dec.}
}

\end{document}